\documentclass[sigconf,table]{acmart} 

\AtBeginDocument{%
  }

\usepackage{multirow}
\usepackage{tabularx}
\usepackage[normalem]{ulem}
\usepackage{url}
\usepackage{placeins}
\usepackage{xspace}
\usepackage{caption}
\usepackage{subcaption}
\usepackage{float}
\usepackage{fancyhdr}
\usepackage{tabularx}
\usepackage{booktabs}
\usepackage{siunitx}
\usepackage{hyperref}
\usepackage{cleveref}
\usepackage{float}
\usepackage{booktabs, multirow}
\hypersetup{
    colorlinks=true,
    linkcolor=blue,
    filecolor=magenta,      
    urlcolor=cyan,
}
\usepackage{enumitem}
\usepackage{booktabs}

\usepackage[utf8]{inputenc}
\usepackage{listings}
\usepackage{dblfloatfix}

\definecolor{backcolor}{rgb}{0.95,0.95,0.92}

\lstdefinestyle{mystyle}{
    backgroundcolor=\color{backcolor},
    basicstyle=\ttfamily\tiny,
    breakatwhitespace=false,
    breaklines=true,
    captionpos=b,
    keepspaces=true,
    showspaces=false,
    showstringspaces=false,
    showtabs=false,
    tabsize=2
}

\usepackage{colortbl}
\definecolor{lightgrayrule}{HTML}{D2D2D2}
\definecolor{rowgray}{HTML}{F1F1F1}

\setlist{leftmargin=3mm}

\definecolor{Author1}{HTML}{e41a1c}
\definecolor{Author2}{HTML}{377eb8}
\definecolor{Author3}{HTML}{ff7f00}
\definecolor{Author4}{HTML}{984ea3}
\definecolor{Author5}{HTML}{ff7f00}

\definecolor{Issue1}{HTML}{e41a1c}
\definecolor{Issue2}{HTML}{377eb8}
\definecolor{Issue3}{HTML}{4daf4a}
\definecolor{Issue4}{HTML}{984ea3}
\definecolor{Issue5}{HTML}{ff7f00}
\definecolor{Issue6}{HTML}{e7298a}
\definecolor{Reviewers}{HTML}{555555}

\newif\ifsubmit
 \submittrue   

\ifsubmit
  \newcommand{\One}[1]{#1}

  \newcommand{\Two}[1]{#1}
  \newcommand{\Three}[1]{#1}
  \newcommand{\Four}[1]{#1}
  \newcommand{\Five}[1]{#1}
  \newcommand{\Six}[1]{#1}

  \newcommand{\OneSide}[1]{}
  \newcommand{\TwoSide}[1]{}
  \newcommand{\ThreeSide}[1]{}
  \newcommand{\FourSide}[1]{}
  \newcommand{\FiveSide}[1]{}

  \newcommand{\OneACSide}[1]{}
  \newcommand{\TwoACSide}[1]{}
  \newcommand{\ROneSide}[1]{}
  \newcommand{\RTwoSide}[1]{}

  \newcommand{\DeletedOne}[1]{}
  \newcommand{\DeletedTwo}[1]{}
  \newcommand{\DeletedTwoo}[1]{}
  \newcommand{\DeletedThree}[1]{}
  \newcommand{\DeletedFour}[1]{}
  \newcommand{\DeletedFive}[1]{}

\else
  \newcommand{\One}[1]{\textcolor{Issue1}{#1}}

   \newcommand{\Two}[1]{\textcolor{Issue2}{#1}}
  \newcommand{\Three}[1]{\textcolor{Issue3}{#1}}
  \newcommand{\Four}[1]{\textcolor{Issue4}{#1}}
  \newcommand{\Five}[1]{\textcolor{Issue5}{#1}}
  \newcommand{\Six}[1]{\textcolor{Issue6}{#1}}

  \newcommand{\OneSide}[1]{\marginpar{\colorbox{Issue1}{\textcolor{white}{\#1}} \textcolor{Issue1}{#1}}}
  \newcommand{\TwoSide}[1]{\marginpar{\colorbox{Issue2}{\textcolor{white}{\#2}} \textcolor{Issue2}{#1}}}
  \newcommand{\ThreeSide}[1]{\marginpar{\colorbox{Issue3}{\textcolor{white}{\#3}} \textcolor{Issue3}{#1}}}
  \newcommand{\FourSide}[1]{\marginpar{\colorbox{Issue4}{\textcolor{white}{\#4}} \textcolor{Issue4}{#1}}}
  \newcommand{\FiveSide}[1]{\marginpar{\colorbox{Issue5}{\textcolor{white}{\#5}} \textcolor{Issue5}{#1}}}

  \newcommand{\OneACSide}[1]{\marginpar{\colorbox{Reviewers}{\textcolor{white}{1AC}} \textcolor{Reviewers}{#1}}}
  \newcommand{\TwoACSide}[1]{\marginpar{\colorbox{Reviewers}{\textcolor{white}{2AC}} \textcolor{Reviewers}{#1}}}
  \newcommand{\ROneSide}[1]{\marginpar{\colorbox{Reviewers}{\textcolor{white}{R1}} \textcolor{Reviewers}{#1}}}
  \newcommand{\RTwoSide}[1]{\marginpar{\colorbox{Reviewers}{\textcolor{white}{R2}} \textcolor{Reviewers}{#1}}}

  \newcommand{\DeletedOne}[1]{\One{\sout{#1}}}
  \newcommand{\DeletedTwo}[1]{\Two{\sout{#1}}}
    \newcommand{\DeletedTwoo}[1]{\Two{\sout{#1}}}
  \newcommand{\DeletedThree}[1]{\Three{\sout{#1}}}
  \newcommand{\DeletedFour}[1]{\Four{\sout{#1}}}
  \newcommand{\DeletedFive}[1]{\Five{\sout{#1}}}
\fi

\newcommand{\system}{MRVS\xspace}
\newcommand{\quotes}[1]{\emph{``#1''}}

\copyrightyear{2026}
\acmYear{2026}
\setcopyright{cc}
\setcctype{by-nc-nd}
\acmConference[CHI '26]{Proceedings of the 2026 CHI Conference on Human Factors in Computing Systems}{April 13--17, 2026}{Barcelona, Spain}
\acmBooktitle{Proceedings of the 2026 CHI Conference on Human Factors in Computing Systems (CHI '26), April 13--17, 2026, Barcelona, Spain}
\acmPrice{}
\acmDOI{10.1145/3772318.3790679}
\acmISBN{979-8-4007-2278-3/2026/04}

\begin{document}


\title[Designing Multi-Robot Ground Video Sensemaking with Public Safety Professionals]{Designing Multi-Robot Ground Video Sensemaking with Public Safety Professionals}



\author{Puqi Zhou}
\email{pzhou@gmu.edu}
\orcid{0000-0002-6486-8883}
\affiliation{%
  \institution{George Mason University}
  \city{Fairfax}
  \state{VA}
  \country{USA}
}

\author{Ali Asgarov}
\email{aliasgarov@vt.edu}
\orcid{0009-0005-6933-4548}
\affiliation{%
  \institution{Virginia Tech}
  \city{Blacksburg}
  \state{VA}
  \country{USA}
}

\author{Aafiya Hussain}
\email{aafiyahussain@vt.edu}
\orcid{0009-0005-8586-9630}
\affiliation{%
  \institution{Virginia Tech}
  \city{Blacksburg}
  \state{VA}
  \country{USA}
}

\author{Wonjoon Park}
\email{wpark814@terpmail.umd.edu}
\orcid{0009-0006-6476-727X}
\affiliation{%
  \institution{University of Maryland}
  \city{College Park}
  \state{MD}
  \country{USA}
}

\author{Amit Paudyal}
\email{apaudya@gmu.edu}
\orcid{}
\affiliation{%
  \institution{George Mason University}
  \city{Fairfax}
  \state{VA}
  \country{USA}
}

\author{Sameep Shrestha}
\email{sshres32@gmu.edu}
\orcid{0009-0009-8732-8802}
\affiliation{%
  \institution{George Mason University}
  \city{Fairfax}
  \state{VA}
  \country{USA}
}

\author{Chia-wei Tang}
\email{cwtang@vt.edu}
\orcid{0009-0005-3384-0668}
\affiliation{%
  \institution{Virginia Tech}
  \city{Blacksburg}
  \state{VA}
  \country{USA}
}

\author{Michael F Lighthiser}
\email{mlighthi@gmu.edu}
\orcid{0009-0009-8614-8834}
\affiliation{%
  \institution{George Mason University}
  \city{Fairfax}
  \state{VA}
  \country{USA}
} 

\author{Michael R Hieb}
\email{mhieb@gmu.edu}
\orcid{0009-0008-1234-1204}
\affiliation{%
  \institution{George Mason University}
  \city{Fairfax}
  \state{VA}
  \country{USA}
} 

\author{Xuesu Xiao}
\email{xiao@gmu.edu}
\orcid{0000-0001-5151-2186}
\affiliation{%
  \institution{George Mason University}
  \city{Fairfax}
  \state{VA}
  \country{USA}
}

\author{Chris Thomas}
\email{christhomas@vt.edu}
\orcid{0000-0002-3226-396X}
\affiliation{%
  \institution{Virginia Tech}
  \city{Blacksburg}
  \state{VA}
  \country{USA}
}

\author{Sungsoo Ray Hong}
\email{shong31@gmu.edu}
\orcid{0000-0001-6050-5404}
\affiliation{%
  \institution{George Mason University}
  \city{Fairfax}
  \state{VA}
  \country{USA}
}

\renewcommand{\shortauthors}{Zhou et al.}


\begin{abstract}
Videos from fleets of ground robots can advance public safety by providing scalable situational awareness and reducing professionals’ burden. Yet little is known about how to design and integrate multi-robot videos into public safety workflows. Collaborating with six police agencies, we examined how such videos could be made practical. In Study 1, we present the first testbed for multi-robot ground video sensemaking. The testbed includes 38 events of interest relevant to public safety, a dataset of 20 robot patrol videos (10 day/night pairs) covering EoI types, and 6 design requirements aimed at improving current video sensemaking practices. In Study 2, we built MRVS, a tool that augments multi-robot patrol video streams with a prompt-engineered video understanding model. Participants reported reduced manual workload and greater confidence with LLM-based explanations, while noting concerns about false alarms and privacy. We conclude with implications for designing future multi-robot video sensemaking tools.
\end{abstract}





\maketitle
\section{Introduction}

\begin{figure*}
  \centering
  \includegraphics[width=1.02\textwidth]{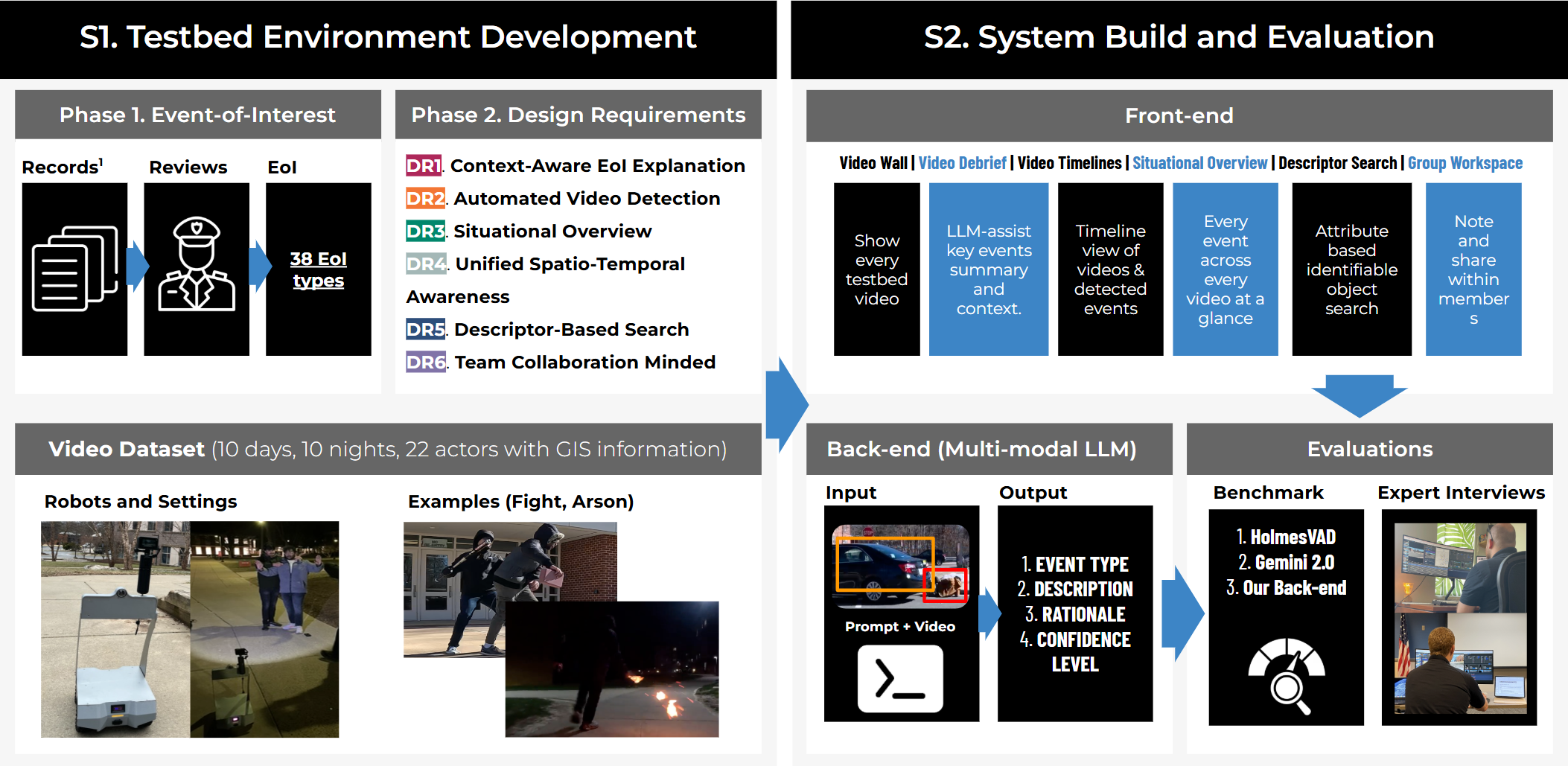}
  \caption{Research flow: (1) Identify 38 EoIs and 6 DRs from records\textsuperscript{1} (13,234 crime records from three US campuses and 10 research anomaly video datasets~\cite{ma2015anomaly, ramachandra2020street, qian2025ucf, wu2020not, wang2020nwpu, zhang2016single, acsintoae2022ubnormal, pranav2020day, abnormal2013lu, rodrigues2020multi}) with five public safety professionals survey reviews and interviews; (2) Create a 20-video multi-robot testbed simulating these EoIs; (3) Build the \system system with front-end interface and multimodal LLM back-end; (4) Evaluate via benchmarking and expert interviews with nine professionals.}
  \vspace{-1em}
  \label{fig:1-workflow}
\end{figure*}
Public safety professionals remain understaffed and disproportionately exposed to injury through in-person operations.
US national data show that public safety professionals experience injury rates more than four times higher than the average occupation~\cite{BLS2014}.
A Department of Justice study of 18 local agencies reported nearly 1,300 injuries in a single year, causing 6,000 missed workdays and about \$2 million in overtime costs~\cite{DOJInjuryStudy}.
Advances in robotics and semi-autonomous control now make fleets of ground robots capable of providing scalable situational awareness for police operations~\cite{liu2025automatic}.
The use of video technologies for situational awareness is not new: agencies have long expanded their reliance on digital imagery, from fixed surveillance cameras~\cite{jenkins2015245} to in-car systems~\cite{reaves2015local}, body-worn cameras~\cite{williams2021body}, and aerial drones~\cite{he2017drone}. 
Most recently, ground robots are being tested in public sector operations~\cite{drew2021multi, bendel2023robots}.
However, early deployments highlight misalignment risks.
The New York Police Department’s K5 patrol robot, for example, was deployed for six months but retired after two months of use due to limited effectiveness and imposed operational burden~\cite{popscik5robot2024}.

While designs that connect public safety professionals with ground-robot video footage promise both societal and technological benefits, there has not been much research in how to make such connections practical and aligned with operational contexts.
Within HCI, public safety professionals have long been studied, but research has focused mainly on fixed surveillance systems~\cite{jenkins2015245, nguyen2016hotspot}, crime prediction~\cite{haque2019exploring}, community engagement~\cite{haque2024we, park2017facilitating}, and decision making~\cite{grgic2019human, veale2018fairness}.
A recurring insight across these studies is the need to involve professionals in the design process~\cite{verma2018confronting}, since interventions developed without their input risk adding burdens rather than support~\cite{tullio2010experience}.
Beyond the public safety domain, HCI research on video interaction has explored video learning~\cite{jin2023collaborative, zhang2025coknowledge}, video editing~\cite{leake2024chunkyedit, huh2025videodiff}, and video assessment~\cite{van2024making, jain2024ai}, showing how users can be supported in extracting insights from large volumes of video data.
Yet despite these advances, video analysis in public safety remains manual and labor-intensive~\cite{tullio2010experience, thacher2008research}.
In short, enabling agencies to make practical use of multi-robot video streams requires substantial adaptation grounded in user-centered design to bridge emerging robotic and AI capabilities with operational needs.

\One{While ground robotics could play a transformative role in high-stakes public-safety operations, we still lack a clear understanding of how these systems should be designed, built, and evaluated in ways that reflect public-safety professionals’ real practices, constraints, and decision-making needs. To mitigate this gap,}
this work first presents a \textbf{testbed environment}\footnote{Link to Testbed: \url{https://github.com/Puqi7/MRVS_VideoSensemaking}} for multi-robot video sensemaking 
\One{, co-designed with experienced public-safety professionals,}
that future researchers can build upon (Study 1, see Fig.~\ref{fig:1-workflow}, S1).
\One{This testbed addresses the current gap by providing a first-of-its-kind design and evaluation environment capable of supporting multi-robot, multi-video workflows in realistic public-safety operational settings.
Next,} this work assesses the testbed's practicality with public safety professionals through the development of an interactive ~\Six{\textbf{Multi-Robot Video Sensemaking System}, }(\textbf{\system}; Study 2, see Fig.~\ref{fig:1-workflow}, S2), conducted in collaboration with six local and state public safety agencies.
\One{\system is the first interactive system that operationalizes multi-robot video sensemaking for public safety by tightly aligning a multimodal AI back-end and a human-centered front-end with public safety professionals' procedures, concerns, and adoption constraints.}

In S1, we (1) identify the types of events captured by ground robots that are relevant to public safety operations (see Fig.~\ref{fig:1-workflow}, Event-of-Interest), (2) construct a multi-robot video dataset with ground-truth labels and context (see Fig.~\ref{fig:1-workflow}, Video Dataset), and (3) derive design requirements for \system-like tools (see Fig.~\ref{fig:1-workflow}, Design Requirements). 
To define \textbf{Events of Interest (EoIs)}, we analyze three years of crime records from three US campuses and 10 public anomaly video datasets in collaboration with five public safety professionals, producing a taxonomy of 38 visually observable events aligned with their operational context.
Building on the EoIs taxonomy, we create a \textbf{Video Dataset} of 10 daytime and 10 nighttime patrol videos, recorded with a ground robot and 22 actors, each with a duration of 22–30 minutes.
Finally, we collaborate with the same five professionals to derive six \textbf{Design Requirements (DRs)} that specify how future systems and interfaces can better support situational awareness with robot-captured video.

In S2, we assess the testbed’s practicality by designing, implementing, and evaluating \system, a \One{novel} system integrating a multimodal LLM back-end and an interactive front-end.
(see Fig.\ref{fig:1-workflow}, Back-end), while the interface provides six core features driven from DRs: Video Wall, Video Debrief, Video Timeline, Situational Overview, Descriptive Search, and Group Workspace (see Fig.\ref{fig:1-workflow}, Front-end).
Benchmark evaluation shows that the back-end achieved F1 score gains of 6\% (day), 23\% (night), and 15\% overall compared to Gemini 2.0~\cite{geminiteam2024geminifamilyhighlycapable}, and substantially outperformed HolmesVAD~\cite{zhang2024holmesvadunbiasedexplainablevideo}. 
To evaluate the front-end, we conduct expert interviews with nine public safety professionals.
They report that \system improved situational awareness, investigation speed, and reduced manual video analysis effort, while also raising concerns about false alarms and privacy risks in \system deployment.
Synthesizing findings from both studies, we outline implications for designing future robot-enabled video sensemaking tools for public safety operations.

This work presents the testbed environment for studying video-sensemaking interfaces generated by multiple ground robots in public safety contexts, enabling future researchers to build and evaluate systems using the expert-validated, labeled video dataset with reproducible design requirements.
\system is the first interactive system grounded in public safety professionals’ practices, providing a reference point for future \system-like systems that leverage multi-robot video workflows for public safety.
\system demonstrates that AI-augmented 
multi-video sensemaking can enable one operator to effectively supervise multiple robots—a critical capability for scaling safety operations under chronic staffing constraints.
This work offers the following contributions:

\begin{itemize}
    \item \textbf{Testbed environment:} We introduce the first multi-robot video sensemaking testbed with (i) a taxonomy of 38 events of interest, (ii) a public ground-robot videos dataset, and (iii) six design requirements derived from public safety professionals.
    \item \textbf{System artifact:} We develop \system, a multi-video sensemaking system with back-end model and front-end interface innovations.
    \item \textbf{Back-end and Front-end Evaluation:} We evaluate back-end through benchmarking against baselines and front-end via expert interviews, providing technical and human-centered insights.
    \item \textbf{Implications for design:} We identify practical, ethical, and privacy considerations in aligning video sensemaking systems with public safety priorities, offering directions for future research.
\end{itemize}
\vspace{-1em}

\section{Related Work}
Gaining situational awareness through ground robots is an emerging problem space.
While few studies have addressed it directly, related work in HCI and Computer Vision (CV) offers useful insights.
In HCI, collaborations with public safety professionals have highlighted the value of user-centered design in safety-critical settings~\cite{calacci2022cop, haque2024we, herrewijnen2024requirements}.
Other HCI studies developed video interaction tools to support sensemaking~\cite{feng2023video2action}, though rarely in public safety contexts.
HCI research has also emphasized ethical and privacy concerns in video surveillance~\cite{corbett2024signs}.
Meanwhile, computer vision has shown longstanding interest in abnormal event detection~\cite{zhang2024holmesvadunbiasedexplainablevideo}, mostly from fixed surveillance perspectives.
Yet the event types examined were not formally elicited from public safety professionals.
In this review, we examine each domain in turn, highlighting both their progress and the gaps that remain.
%

\vspace{0.5em}
\textit{\textbf{HCI: Working with Public Safety Professionals.}} 
When introducing technological aids, including AI-driven tools, in safety-critical contexts, grounding design in the situated practices of public safety professionals is critical. 
Prior work highlights a persistent misalignment between sophisticated video technologies and interpretation expertise, underscoring the need for systems that augment rather than replace professional judgment~\cite{tullio2010experience, thacher2008research}. 
To be adopted, such tools must integrate seamlessly into existing workflows rather than impose additional burdens~\cite{hall2022explainable}. 
Studies show that officers value explainability framed in policing terms~\cite{herrewijnen2024requirements, deeb2019understanding} and moderate forms of AI assistance that reduce cognitive load while maintaining professional control~\cite{bayley1984learning, willis2018improving}. 
Adoption is shaped by organizational dynamics and stakeholder engagement~\cite{kawakami2024studying}, reinforcing the importance of trust~\cite{saxena2025ai}, accessibility~\cite{balasubramaniam2024bridging}, and institutional fit~\cite{veale2018fairness}. 

Although ground robots are emerging in public safety, no HCI study has examined this space; related work focuses on other domains. 
Agencies have expanded sociotechnical infrastructures with diverse video sources—from fixed CCTV~\cite{zhang2019edge} to body cameras~\cite{williams2021body}, drones~\cite{he2017drone}, and automated analytics~\cite{reaves2015local, bendel2023robots}, yet empirical understanding of how these tools reshape frontline policing remains limited~\cite{brayne2017big}. 
Research shows surveillance technologies often increase, rather than alleviate, officers’ cognitive burden when monitoring massive, fragmented streams, while data-driven approaches overlook experiential knowledge essential in high-stakes contexts~\cite{nguyen2016hotspot, verma2018confronting}.
While AI systems have been piloted in areas such as patrol planning~\cite{veale2018fairness}, crime forecasting~\cite{simmons2016quantifying}, and crime mapping~\cite{haque2019exploring}, questions remain about aligning emerging multi-robot videos with frontline professionals' sensemaking practices.
%

\vspace{0.5em}
\textit{\textbf{HCI: Human-Video Interaction.}} 
Extensive research has advanced interactive designs for video streams and content, yet public safety investigations still rely on manual, labor-intensive practices. 
Many systems support video sensemaking, including tools for navigation~\cite{yu2010multiplayer}, retrieval~\cite{tran2023integrating}, and collaborative analysis across multiple streams~\cite{shipman2003generation, xia2014exploring}.  
Recent work has explored chunk-based editing of interview footage~\cite{leake2024chunkyedit}, instruction-following video interactions~\cite{feng2023video2action}, and crowd-assisted annotation~\cite{kim2014crowdsourcing}. 
Systems such as ChronoViz~\cite{fouse2011chronoviz} and CrossA11y~\cite{liu2022crossa11y} indicate that aligning notes, metadata, and multimodal cues deepens interpretation. 
Other studies highlight the heavy cognitive demands of reviewing complex video~\cite{thalmann2019does, liu2018Concept}, challenges of assessing video credibility~\cite{hughes2024viblio}, and interface designs for multi-video layouts~\cite{alkhudaidi2025perceptions, nguyen2016hotspot}.  
Despite this progress, design assumptions behind these systems are often rooted in creative or interaction settings rather than policing, leaving challenges of fragmented, distributed multi-robot video largely unexplored.
%
\begin{table*}[!b]
\vspace{1em}
\centering
{\small
\begin{tabularx}{\textwidth}{l l p{5cm} c X}
\toprule
\textbf{PID} & \textbf{Rank} & \textbf{Agency} & \textbf{Years Exp.} & \textbf{Primary Focus Areas} \\
\midrule
P1 & Lieutenant & Arlington County Police Department & 22 & Real-time crime, canine, SWAT, drones, civil disturbance \\
P2 & Captain & City of Fairfax Police Department & 18 & Patrol, criminal investigation, community service \\
P3 & Detective & Manassas City Police Department & 8 & Post-incident investigation, drones \\
P4 & Detective & George Mason University Police Department & 7 & Post-incident investigation \\
P5 & Sergeant & Virginia State Police Department & 15 & Real-time crime, post-incident investigation, drones \\
\bottomrule
\end{tabularx}
}
\caption{Study 1 Participant Profiles.}
\label{tab:participant-s1}
\end{table*}

\vspace{0.5em}
\textit{\textbf{HCI: Ethical and Privacy Considerations.}} HCI research has examined the ethical and privacy implications of surveillance, including community perceptions of being monitored, when adopting intelligent, AI-driven video analysis for public safety. 
Studies show that while robotic patrols and AI detection tools promise safer engagements~\cite{lakecharles_robotdog, marcu2023would}, they risk harm when introduced without transparency or oversight~\cite{petty2018reclaiming, shin2024delivering}. 
Public backlash to San Diego’s covert streetlight program illustrates how hidden monitoring undermines trust and fuels community demands for accountability~\cite {whitney2021hci}. 
Research notes that video capture in outdoor public environments is often considered acceptable~\cite{yao2017free}, but only when safeguards are in place to balance safety benefits with privacy protection~\cite{benton2023using}. 
%

\vspace{0.5em}
\textit{\textbf{CV: Abnormality Detection.}} 
The computer vision community has extensively developed video understanding models, classifiers, and datasets for public safety, often framed as ``anomaly detection''.
However, most efforts focus on fixed-perspective surveillance footage rather than ground-robot video, and event types have not been defined in consultation with public safety professionals, limiting their value as practical benchmarks.
Advanced vision models~\cite{Wangetal2018, Zhangetal2023} trained on fixed video perform poorly on noisy, mobile streams.
Core challenges include context-dependent anomaly definitions, data imbalance, and real-time constraints~\cite{Georgescuetal2020, Honnegowdaetal2024}. 
Methods range from unsupervised normality learning~\cite{Wangetal2020}, to weakly-supervised video-level MIL~\cite{Sultani_2018_CVPR, Sunetal2023}, and \textit{supervised} frame-level labeling~\cite{Maetal2025}, with advanced such as autoencoder reconstruction~\cite{Hasanetal2016}, frame prediction~\cite{Liuetal2017}, adversarial training~\cite{Royetal2020}, memory models~\cite{Ahnetal2025}, and transformers for temporal reasoning~\cite{Yangetal2023}. 
Recent work explores recurrent autoencoders~\cite{Wangetal2022} and vision-language models such as Video-LLaMA2~\cite{cheng2024videollama2advancingspatialtemporal} and Gemini~\cite{ geminiteam2024geminifamilyhighlycapable}, enhancing scene reasoning while raising efficiency and deployment challenges~\cite{Xuetal2022, muralidharan2024enhanced}.
Existing abnormal resources include campus footage~\cite{wang2020nwpu}, online clips~\cite{qian2025ucf}, or long-form surveillance video~\cite{pranav2020day}.

As we reviewed the four domains, we found that each offers useful directions for designing practical tools in multi-robot video sensemaking, yet key gaps remain.
First, no design or artifact research has addressed how public safety professionals might interact with multiple video streams from the moving perspectives of ground robots.
Second, event types relevant to real-world public safety operations have not been systematically defined in collaboration with professionals.
Third, without such event definitions, no datasets exist to simulate multi-robot video sensemaking scenarios for research and design.
Addressing these gaps requires a testbed environment to support the development and evaluation of new tools in the multi-robot video sensemaking problem space. 
\section{S1. Formative Study}
\label{sec:formative-results}
We aim to build a testbed environment grounded in the operational realities of public safety work and identify professionals’ design requirements for future \system-like systems development.
We conducted a two-phase formative study to ground our design in the realities of public safety practice with five public safety professionals combining a survey (Phase~1) and expert interviews (Phase~2) to anchor our design in real-world practice.
All study procedures were reviewed and approved by our university’s IRB, and informed consent was obtained from all participants.
Specifically, this study addressed the following~\Six{two} research questions:
\begin{itemize}
    \item {\textbf{RQ1.}} Which events captured by multi-robot video are perceived as relevant and urgent for public safety? 
    \item {\textbf{RQ2.}} How do public safety professionals conduct video investigations with technological aids, what challenges remain, and how do they collaborate during the process?
\end{itemize}
\noindent{}Beyond the two RQs, we explored professionals' perceptions of multi-robot video sensemaking as an emerging technology, what societal and ethical considerations arise in its deployment, and what further issues future researchers should keep in mind when engaging with this domain.

\subsection{Participants and Recruitment}
Police work is high-stakes, time-sensitive, and bound by strict confidentiality and protocols, making direct recruitment for our studies challenging. 
To apply a human-centered design approach to building \system into public safety workflow, we first established trust and long-term engagement with police agencies. 
Over six months, three authors presented the \system vision seven times across multiple police departments in one US state, gradually building relationships with three representatives from different agencies.
These representatives agreed to support recruiting other police officers for future interviews and system evaluations.
To facilitate recruitment, we created an outreach package including a short explanatory video, a slide deck outlining the project scope, and a one-page summary.
Agency representatives circulated these materials internally to ensure confidentiality and organizational compliance.
Participants completed a screening survey capturing gender, department, role, video-investigation experience, and years of service.
This trust-based recruitment approach yielded five active-duty participants spanning patrol, dispatch, investigation, and supervisory roles, with 7–22 years of experience (see Table~\ref{tab:participant-s1}).
All five participants completed both Phase~1 and Phase~2.  
%
\begingroup
\setlength{\dbltextfloatsep}{3pt}  
\setlength{\dblfloatsep}{6pt}      
\setlength{\abovecaptionskip}{2pt}
\setlength{\belowcaptionskip}{2pt}

\subsection{Phase 1. Events of Interest (EoIs)} 
Phase~1 addressed RQ1 through an online survey designed to identify EoIs relevant to \system multi-robot video public safety scenarios and the collection of contextual resources.
\subsubsection{Methodology}
\Two{We constructed an online EoI survey instrument (see supplement) and distributed it to five public-safety professionals to obtain a practitioner-grounded set of EoIs (Table~\ref{tab:EoIfinal}).
Our EoI derivation process consisted of three steps: (1) preparing the set of potential EoIs, (2) gathering public-safety professionals’ feedback on their validity and priority, and (3) analyzing the results.}

\textbf{\Two{Step 1. EoIs Preparation:}} We first constructed a candidate set of EoIs balancing operational realism, practical relevance, and reduced participant burden through a three-step process:
(1) \textit{Data collection}: We assembled an initial pool of candidate records from two sources: (a) 13,234 daily crime logs collected over three years across three universities in one of the U.S. states and (b) ten public research video datasets on anomaly detection and crime video analysis~\cite{ma2015anomaly, ramachandra2020street, qian2025ucf, wu2020not, wang2020nwpu, zhang2016single, acsintoae2022ubnormal, pranav2020day, abnormal2013lu, rodrigues2020multi}.
\Two{For the daily crime logs, we extracted the existing fields \textit{Nature of Case}, \textit{Brief Description}, and \textit{Offenses}.
For the video datasets, we collected available \textit{event types} and \textit{descriptions}; when unavailable, two authors independently reviewed videos, drafted descriptions, and resolved differences through consensus coding~\cite{strauss1987qualitative} to complete our initial record archive.
(2) \textit{Filtering records}: Using descriptions from both sources, we removed records that (a) could not be visually detected from robot-captured footage or (b) would be impractical to simulate (e.g., sexual offenses).
(3) \textit{Classification}: With input from our public-safety co-author, we reviewed all remaining records and defined event types aligned with U.S. public-safety standards.
Category naming drew from the Clery Act~\cite{cleryact}, a legal standard for campus-crime classification, and event types used in existing abnormal-event video datasets.}
\Two{To categorize the 40 candidate events of interest, five authors conducted an in-person collaborative affinity diagramming session~\cite{lucero2015using}.
Authors iteratively clustered EoIs and developed high-level themes through synchronous discussion, informed by a coauthor’s public safety domain expertise. Each EoI was assigned only after unanimous consensus, followed by secondary review to ensure themes were mutually exclusive and differentiated. This produced six categories: (a) Property and Environmental Incidents, (b) Public Order Disturbances, (c) Vehicle and Mobility Incidents, (d) Suspicious or Unusual Behavior, (e) High-Risk Threats, and (f) Miscellaneous Activities. Finally, taxonomy was refined with concise definitions and representative visual cues to ground the survey.
Each category was accompanied by concise definitions and representative visual cues derived during data collection, with example images sourced from public materials, establishing a structured foundation for the survey.}

\textbf{\Two{Step 2. EoIs Survey:}}
With this survey instrument prepared, we invited public safety professionals to complete the online EoI survey via email. 
\Two{Participants first viewed the six category definitions with example events and representative visual cues, then identified any missing event types within each category in open text to refine the taxonomy.
For each of the 40 candidate EoIs, participants answered three questions: (1) ``Is this event relevant to public safety?'' (binary), (2) ``How important do you consider this event for public safety?'' (7-point Likert), and (3) ``How urgent do you consider this event for public safety?'' (7-point Likert).
These questions were designed to assess each EoI’s relevance to our problem space and to capture perceived urgency and importance to establish EoI priority.
}

\Two{\textbf{Step 3. EoIs Analysis:} 
For each candidate EoI, we calculated the proportion of officers who judged it as relevant, the mean ($\text{M}$) and standard deviation ($\text{Std}$) of both urgency and importance ratings.
Analysis proceeded in three steps:}
\Two{(1) excluding events that were either unanimously deemed irrelevant or judged relevant by only one officer with low urgency and importance ratings ($\text{M}<2$, $\text{Std}<0.5$);} 
\Two{(2) applying $k$-means clustering to group the remaining events by their mean urgency and importance ratings with $k{=}4$ selected via an elbow analysis~\cite{thorndike1953};}
\Two{and (3) adding events from open-ended responses, defining priorities through Phase~2 expert interviews.}
\subsubsection{Results}
The Phase 1 survey produced a final taxonomy of 38 EoI types (see Table~\ref{tab:EoIfinal}), grouped into four operational levels: \textit{emergency, urgent, moderate, advisory}.
\Two{Eight candidate events were excluded: six unanimously judged irrelevant and two judged relevant by only one officer with consistently low urgency and importance ratings (e.g., \textit{Climbing fence}).
Six additional events emerged from open-ended responses, including \textit{Incident exposure} and \textit{Propping open doors}, capturing context-specific concerns beyond predefined survey EoIs.}
\Two{K-means clustering revealed a clear boundary between response-critical events and routine observations: \textit{Emergency} events showed the strongest agreement, with the highest mean ratings with relatively low dispersion ($\text{M}_{\text{Imp}} \approx 5.84$; $\text{Std}_{\text{Imp}} \approx \mathbf{1.64}$; $\text{M}_{\text{Urg}} \approx 6.27$; $\text{Std}_{\text{Urg}} \approx \mathbf{1.79}$), indicating consensus on immediate threats, whereas disagreement concentrated in the \textit{Moderate} presenting the largest variance across both dimensions ($\text{Std}_{\text{Imp}} = \mathbf{2.39}$; $\text{Std}_{\text{Urg}} = \mathbf{2.88}$), suggesting triage sensitivity to situational context.
Based on these, police co-authors proposed the final four-level taxonomy to reflect operational prioritization.
This four-level structure served as the foundation for Phase 2, where expert interviews successfully validated these clustering levels and enriched them with additional metadata (e.g., entity categories).}
\renewcommand\tabularxcolumn[1]{m{#1}} 
\begin{table*}[!t]
\centering

\begin{tabularx}{\textwidth}{rXcclrXccl}
\arrayrulecolor{black}
\specialrule{.1em}{0pt}{0pt}
\rowcolor{black}
{\color{white}\textbf{No.}} &
{\color{white}\textbf{Event}} &
{\color{white}\textbf{Priority}} &
{\color{white}\textbf{Occur.}} &
{\color{white}\textbf{Label}} &
{\color{white}\textbf{No.}} &
{\color{white}\textbf{Event}} &
{\color{white}\textbf{Priority}} &
{\color{white}\textbf{Occur.}} &
{\color{white}\textbf{Label}} \\
\specialrule{.1em}{0pt}{0pt}

\arrayrulecolor{lightgrayrule}

{\footnotesize 1}  & {\footnotesize Arson} & {\footnotesize Emergency} & {\footnotesize 3}  & {\footnotesize Crime} & {\footnotesize 20} & {\footnotesize Trespassing} & {\footnotesize Urgent}   & {\footnotesize 0}  & {\footnotesize Crime} \\ \midrule
{\footnotesize 2}  & {\footnotesize Burglary} & {\footnotesize Emergency} & {\footnotesize 0}  & {\footnotesize Crime} & {\footnotesize 21} & {\footnotesize Drunkenness} & {\footnotesize Urgent}   & {\footnotesize 10}  & {\footnotesize Civil} \\ \midrule
{\footnotesize 3}  & {\footnotesize Robbery} & {\footnotesize Emergency} & {\footnotesize 17} & {\footnotesize Crime} & {\footnotesize 22} & {\footnotesize Snatching Bag} & {\footnotesize Urgent}   & {\footnotesize 4} & {\footnotesize Civil} \\ \midrule
{\footnotesize 4}  & {\footnotesize Assault} & {\footnotesize Emergency} & {\footnotesize 13} & {\footnotesize Crime} & {\footnotesize 23} & {\footnotesize Bag Left Behind} & {\footnotesize Urgent}   & {\footnotesize 1}  & {\footnotesize Civil} \\ \midrule
{\footnotesize 5}  & {\footnotesize Shooting} & {\footnotesize Emergency} & {\footnotesize 1}  & {\footnotesize Crime} & {\footnotesize 24} & {\footnotesize Indecent Exposure} & {\footnotesize Urgent}   & {\footnotesize 0}  & {\footnotesize Civil} \\ \midrule
{\footnotesize 6}  & {\footnotesize Explosion} & {\footnotesize Emergency} & {\footnotesize 0}  & {\footnotesize Crime} & {\footnotesize 25} & {\footnotesize Unattended Domestic Animals} & {\footnotesize Urgent} & {\footnotesize 3}  & {\footnotesize Civil} \\ \midrule
{\footnotesize 7}  & {\footnotesize Kidnapping} & {\footnotesize Emergency} & {\footnotesize 4}  & {\footnotesize Crime} & {\footnotesize 26} & {\footnotesize Medical Emergencies} & {\footnotesize Urgent} & {\footnotesize 2}  & {\footnotesize Civil} \\ \midrule
{\footnotesize 8}  & {\footnotesize Weapon Holding} & {\footnotesize Emergency} & {\footnotesize 4}  & {\footnotesize Crime} & {\footnotesize 27} & {\footnotesize Illegal Parking} & {\footnotesize Moderate} & {\footnotesize 3}  & {\footnotesize Civil} \\ \midrule
{\footnotesize 9}  & {\footnotesize Destruction/Damage/ Vandalism} & {\footnotesize   ~~~Urgent} & {\footnotesize 9} & {\footnotesize Crime} & {\footnotesize 28} & {\footnotesize People Falling} & {\footnotesize Moderate} & {\footnotesize 11}  & {\footnotesize Civil} \\ \midrule
{\footnotesize 10} & {\footnotesize Theft from Vehicle} & {\footnotesize Urgent} & {\footnotesize 6}  & {\footnotesize Crime} & {\footnotesize 29} & {\footnotesize Person Smoking} & {\footnotesize Moderate} & {\footnotesize 3}  & {\footnotesize Civil} \\ \midrule
{\footnotesize 11} & {\footnotesize Theft from Building} & {\footnotesize Urgent} & {\footnotesize 2}  & {\footnotesize Crime} & {\footnotesize 30} & {\footnotesize Prohibited U-turns} & {\footnotesize Moderate} & {\footnotesize 2}  & {\footnotesize Civil} \\ \midrule
{\footnotesize 12} & {\footnotesize Motor Vehicle Theft} & {\footnotesize Urgent} & {\footnotesize 2}  & {\footnotesize Crime} & {\footnotesize 31} & {\footnotesize Jaywalking} & {\footnotesize Moderate} & {\footnotesize 8}  & {\footnotesize Civil} \\ \midrule
{\footnotesize 13} & {\footnotesize Abuse} & {\footnotesize Urgent} & {\footnotesize 2}  & {\footnotesize Crime} & {\footnotesize 32} & {\footnotesize Cars Stopping on Road} & {\footnotesize Moderate} & {\footnotesize 3}  & {\footnotesize Civil} \\ \midrule
{\footnotesize 14} & {\footnotesize Brawling} & {\footnotesize Urgent} & {\footnotesize 13} & {\footnotesize Crime} & {\footnotesize 33} & {\footnotesize Harassment/Stalking} & {\footnotesize Moderate} & {\footnotesize 3}  & {\footnotesize Crime} \\ \midrule
{\footnotesize 15} & {\footnotesize Crowds Escaping} & {\footnotesize Urgent} & {\footnotesize 0}  & {\footnotesize Civil} & {\footnotesize 34} & {\footnotesize Loitering} & {\footnotesize Advisory} & {\footnotesize 2}  & {\footnotesize Civil} \\ \midrule
{\footnotesize 16} & {\footnotesize Obstructing Justice} & {\footnotesize Urgent} & {\footnotesize 0}  & {\footnotesize Crime} & {\footnotesize 35} & {\footnotesize Crowd Gathering} & {\footnotesize Advisory} & {\footnotesize 1}  & {\footnotesize Civil} \\ \midrule
{\footnotesize 17} & {\footnotesize Carrying Suspicious Object} & {\footnotesize Urgent} & {\footnotesize 5} & {\footnotesize Civil} & {\footnotesize 36} & {\footnotesize Wrong-way Driving} & {\footnotesize Advisory} & {\footnotesize 2}  & {\footnotesize Civil} \\ \midrule
{\footnotesize 18} & {\footnotesize Hit and Run} & {\footnotesize Urgent} & {\footnotesize 7} & {\footnotesize Crime} & {\footnotesize 37} & {\footnotesize Wearing Face Mask} & {\footnotesize Advisory} & {\footnotesize 3}  & {\footnotesize Civil} \\ \midrule
{\footnotesize 19} & {\footnotesize Road Accidents} & {\footnotesize Urgent} & {\footnotesize 6}  & {\footnotesize Civil} & {\footnotesize 38} & {\footnotesize Propping Doors Open} & {\footnotesize Advisory} & {\footnotesize 1}  & {\footnotesize Civil} \\

\arrayrulecolor{black}
\specialrule{.1em}{0pt}{0pt}
\end{tabularx}
\caption{Events Classification with Occurrence Count in the testbed environment}

\label{tab:EoIfinal}
\end{table*}
\subsection{Phase 2. Design Requirements (DRs)}
Phase~2 addressed RQ2 through in-person semi-structured interviews~\cite{lazar2017research} with the same five participants. 
The interviews produced six DRs for designing tools that enhance public safety professionals’ current video sensemaking practices by enabling the use of multi-robot video sensemaking capabilities.

\subsubsection{Methodology}
Each interview followed a semi-structured format designed to ensure consistency and elicit participants’ perspectives on current practices and potential system implications. 
A prepared slide deck guided the discussions across three themes.
The first revisited Phase~1 results, prompting reflections on the EoIs survey, event types deemed irrelevant, and types of professionals believed ground-robot footage should capture more extensively. 
The second focused on video investigation practices, addressing current technology use, challenges faced, desired functionalities, and visual cues professionals rely on to identify people or objects. 
The third examined system implications, exploring concerns about adopting MRVS-like capabilities, anticipated benefits, and suggestions for future development.
Each sessions lasted 38–65 minutes, were audio-recorded with consent, automatically transcribed, manually reviewed, and corrected by the lead author for accuracy.

We analyzed transcripts using Braun and Clarke’s thematic analysis framework~\cite{braun2006using}, informed by Saldaña’s coding methodology~\cite{saldana2015coding}. 
The lead and fourth authors independently conducted open coding across all transcripts. 
To ensure analytical rigor, we assessed inter-rater reliability on the pre-reconciliation codes, achieving substantial agreement (Cohen’s $\kappa$ = 0.72). The authors then met to discuss discrepancies and iteratively refined and consolidated codes into higher-level themes, producing a finalized codebook. These themes were synthesized into six DRs specifying how \system should support public safety video work.

\subsubsection{Results}
We developed seven interrelated themes illustrating how public safety professionals engage with multiple videos in practice (T1-\Two{T7}). 
Across themes (T1-T6) surface tensions between frontline needs and existing technological support.
We present each theme alongside its corresponding design requirement (DR1–DR6) and one additional insightful design consideration.
\label{sec:eoi-definition}
\paragraph{T1. \textbf{Context-Sensitive Recipes for Event Detection.}}
\label{sec:T1}
While detecting an event type such as a parking violation may appear straightforward, the ``recipe'' for defining and justifying an event is highly context-dependent, hinging not only on the immediate situation but also on agency policies, spatial constraints, and operational norms.
For example, a parked vehicle might be a delivery or a crime signal in preparation (P2).
Likewise, gatherings acceptable in public may be treated as suspicious loitering on private property (P4).
Professionals emphasized that EoIs shift with organizational capacity and priorities; the same event may draw attention in one agency but not another, depending on staffing and coverage (P1-P3).
EoIs are also distinguished between real-time alerts and post-incident review~\Two{(N=3/5)} :
for instance, \quotes{Skateboarding isn’t alert-worthy, but it matters later} (P3), and \quotes{Parking needs no response, but we want the data} (P5).
These reflections highlight that event significance is inseparable from institutional and temporal context. 
Accordingly, professionals expect video sensemaking systems to move beyond simple flagging by explaining why behavior is abnormal (P2, P3) and simplifying cognitively heavy taxonomies into clearer groupings such as \quotes{vehicle, people, others} (P4), for clearer criminal-civil distinctions, adopting to finalize our EoIs results (details see Table~\ref{tab:EoIfinal}).
\Six{\noindent\textbf{DR1. Context-Aware EoIs Explanation.} 
\system should support flexible, contextual EoIs tailored to agency needs, with transparent
AI reasoning and explanations.}
\paragraph{T2. \textbf{The \textit{Grind} of Single-Video Analysis.}}
\label{sec:T2}
Investigating a single video remains one of the most cognitively exhausting and time-consuming tasks in public safety work, revealing the need for more effective ways to navigate and interpret footage.  
Public safety professionals described spending hours manually scanning through footage to locate mere seconds of relevant content, typically fast-forwarding at 8x or 16x speed~\Two{(N=4/5)}. 
This requires sustained focus and risks missing subtle cues, especially in low-light or low-resolution footage. As P3 put it, \quotes{I could be looking for 15 seconds of video in 3 or 4 hours of footage… I can’t blink, or I’ll miss it}.  
Agencies sometimes delegate this labor to junior staff to preserve the senior staff’s time (P1), yet chronic understaffing still leads to overload and leaves lower-urgency cases backlogged (P3, P5). 
Technical flaws further undermine trust: misaligned timestamps and interfaces lacking frame-accurate scrubbing make review difficult and not useful in their workflow~\Two{(N=3/5)}.
As P4 remarked, \quotes{I’ve seen videos say 2 AM when it’s clearly daylight—how can we use that in court?}  
Participants expressed cautious interest in AI tools, stressing that decision-making must remain human-controlled. 
They sought support that surfaces key moments with a transparent rationale. 
As P2 emphasized, \quotes{Quick detection is not enough. I need to know why it thinks this is evidence.} 
Concerns include unreliable outputs creating additional verification work and over-reliance on unexplainable results that could undermine legal admissibility (P1, P4).
Professionals valued the small and accurate assistance over automation, which may simply shift verification burdens onto them~\Two{(N=5/5)}.\\
\Six{\noindent\textbf{DR2. Automated Video Detection.}
\system should enable efficient review through automated abnormal event summaries, AI reasoning, and timeline navigation to reduce manual effort.}
\paragraph{T3. \textbf{The \textit{Juggle} of Multi-Video Sensemaking}. }
\label{sec:T3}
Making sense of multiple video streams compounds the difficulty of situational awareness, as professionals must monitor several feeds simultaneously while extracting and connecting dispersed events into a coherent picture.
While additional cameras provided wider coverage and multiple angles, participants described the experience of juggling them as cognitively overwhelming~\Two{(N=4/5)}.
They struggled to decide which screen deserved attention, often missing important activity elsewhere, and found it difficult to piece together a continuous storyline across feeds in real time (P2, P5).
As P3 explained, \quotes{I can pull four videos to watch at once, but I can’t fully make sense of each one}.
The fragmented video sources came from different platforms, forcing professionals to switch between systems and disrupting workflow~\Two{(N=2/5)}. 
As P1 noted, \quotes{We have someone monitoring calls, pulling up whatever camera they can, but it’s all disconnected. You have to jump system to system, with no single place showing everything that matters}. 
Participants envisioned tools that could relieve this “juggling” by moving beyond raw video display toward an integrated, event-centric overview. Similar to a police report, which has a high-level overview and recorded anomalies with rank urgency.
As P3 summarized, \quotes{Help me see everything at a glance. Flag what’s abnormal, show me what’s important first}.  
\Six{\noindent\textbf{DR3. Situational Overview.}
\system should offer a unified overview across video streams for event comparison, urgency prioritization, and anomaly detection to support rapid decisions.}
\paragraph{T4. \textbf{When Space and Time Collide}.}
\label{sec:T4}
Video investigation is hindered by the collision of space and time, as professionals must determine which sources captured a given location and whether the footage is still accessible (P3, P4). 
This tension between spatial coverage and temporal accessibility creates a substantial burden, forcing them to manually reconstruct whether useful video exists.
Current infrastructures are fragmented: traffic cameras offer real-time monitoring but often \quotes{don’t record} (P4); private business footage, though critical, can take days or weeks if owners are unavailable (P2).
Even when accessible, fixed cameras leave blind spots in mid-blocks, trails, and rural roads, making it unclear whether an incident was ever captured~\Two{(N=3/5)}. 
Professionals emphasized that the first step in post-incident work is simply confirming whether useful video exists, a process that requires manually scanning across six to seven disconnected systems—body-worn, in-car, drone, traffic, and private feeds—all with separate logins and inconsistent retention. 
This verification process is time-consuming and cognitively demanding, often complicated by vague reporting, such as \quotes{a business was broken into overnight (P2).} 
Participants envisioned that instead of piecing together coverage maps, an integrated platform could synchronize all video sources onto a unified map and timeline views. 
Such a system would help professionals to quickly identify where and when coverage is available, close geographic gaps, and streamline the reconstruction of movement timelines.
\Six{\noindent\textbf{DR4. Unified Spatio-Temporal Awareness.}
\system should synchronize fragmented feeds in an interface combining map and
timeline views for seamless verification and reconstruction.}
\paragraph{T5. \textbf{Attribute Without a Place to Search}.} 
\label{sec:T5}
Even when professionals know which attributes of a person or vehicle they want to locate, current systems can not operationalize those descriptors, forcing long manual video scanning. 
Rather than depending on faces or license plates, they look for relatively stable attributes such as hair, clothing, pants, shoes, or vehicle features like model, make, color, and visible damage~\Two{(N=5/5)}. 
As P1 emphasized, \quotes{People change jackets easily even after minor crime, but they don’t usually change shoes.} 
Similarly, P3 noted, \quotes{Nine times out of ten, I find someone by their pants or shoes, not their face.} 
Yet these clues are often vague or incomplete.
A witness may only recall a red hoodie, leading to hours of scanning across feeds.
Vehicle tracking presents comparable challenges: plates are easily obscured or swapped, making color, model, and make more reliable identifiers (P2).
Despite their importance, current video systems cannot link descriptors across sources, leaving professionals with labor-intensive searches~\Two{(N=3/5)}.
Participants emphasized the need for descriptor-driven search that connects attributes across footage and provides transparent explanations of which features were matched and why. 
Such functionality would improve efficiency and align with policy constraints and public concerns, limiting facial recognition (N=3/5).
\Six{\noindent\textbf{DR5. Descriptor-Based Search.}
\system should support search for people and vehicles using appearance attributes to locate entities across video data quickly.}
\begin{figure*}
  \centering
  \vspace{1em}
  \includegraphics[width=1\textwidth]{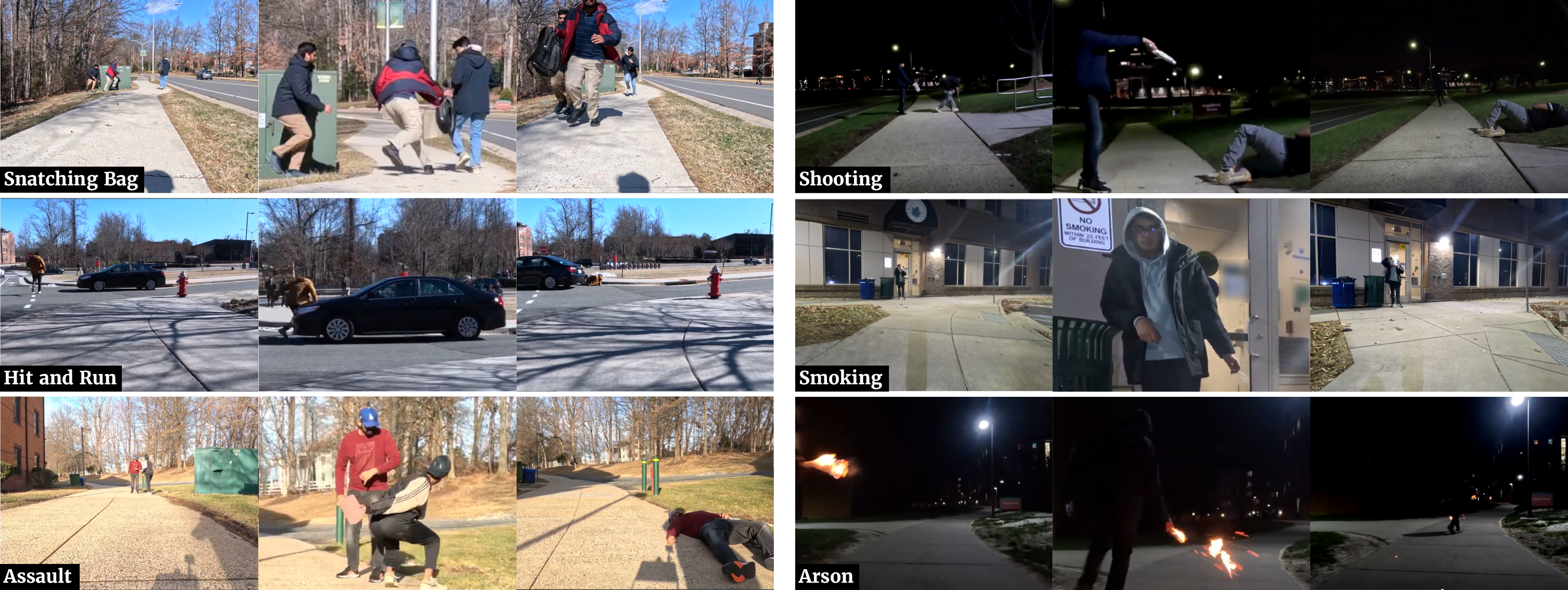}
  \caption{Examples of different anomalies in our testbed shown in sequences. Each second column is manually zoomed in.}
  \label{fig:dataset}
\end{figure*}
\paragraph{T6. \textbf{Gaps in Collaborative Sensemaking}.}
\label{sec:T6}
Teams struggle to coordinate their findings and share interpretations across video analysis tasks, underscoring the need for better collaborative support.
Currently, though video investigations are highly collaborative, often spanning multiple teams, shifts, and extended periods, there is little support for persisting progress, sharing annotations, or enabling seamless handovers between public safety professionals.
Critical updates were typically shared informally through texts, emails, or verbal briefings, which were prone to being lost, misunderstood, or overlooked~\Two{(N=2/5)}. 
P3 noted \quotes{We just email clips around and hope the next person watches the right part.}
This lack of continuity resulted in duplicate work, missed leads, and inconsistent conclusions, particularly during long investigations or shift changes.
The professionals expressed frustration at the need to manually track timestamps, notes, and clip links, which then had to be transferred into emails or printable formats to share with the team (P4). 
Such fragmented workflows created unnecessary barriers in situations where speed, accuracy, and team coordination were critical.
These findings highlight the need for systems facilitating team-based workflows by supporting persistent annotations, shared workspaces, and cross-shift progress tracking, ensuring the continuity and collective sensemaking throughout investigations.\\
\Six{\noindent\textbf{DR6. Team Collaboration Minded.}
Building on T6, \system should facilitate collaborative annotation, validation, and event management to
support workflows and continuity across shifts.}
\paragraph{\Two{T7. \textbf{Socio-Technical Preconditions for Patrol Robot Scenario.}}}
\Two{
Professionals recognized patrol robots' potential as force multipliers for police work, noting their deterrent effect similar to parking marked cars in high-crime areas. However, they emphasized that successful deployment requires more than just technical readiness.
Key concerns included public acceptance and privacy~\Two{(N=4/5)}. 
Participants stressed the need for transparency, clear operational boundaries, and context-appropriate privacy protections with robots minimizing identifiable capture during routine patrol while allowing richer footage during authorized investigations (P2). 
Several noted that while police-styled robots could enhance deterrence
\quotes{Knowing the robot is patrolling here will disperse some people. You could claim that as a win (P4)}
, preventing them from becoming attack targets in reality needs to be carefully considered.
Practical deployment challenges centered on cost sustainability, hardware robustness, and reliable performance in rough terrain. P1 noted: \quotes{Budget's a huge thing and changes every year—you have to make cuts.}
\textbf{Design Consideration}: Ensure patrol robot deployment respects community norms and institutional capacity through transparency logs, operational boundaries, and durable infrastructure that balances effectiveness with public trust.}

\section{Video Dataset}
We present a first-of-its-kind video dataset\footnote{Link to Dataset: \url{https://huggingface.co/datasets/Puqi7/MRVS_anomaly_long_video_dataset}} captured by a patrolling ground robot to establish a testbed environment for multi-robot video sensemaking.
Unlike existing anomaly video datasets, ours is grounded in co-defined EoIs with public safety professionals, enacted in real-world environments, and recorded under police-guided patrol scripts specifying locations and event types.
This dataset offers a realistic benchmark for evaluating future systems leveraging robot-captured video in public safety.
Representative scripted EoIs are shown in Fig.~\ref{fig:dataset}.

Grounded in the refined EoI types in Table~\ref{sec:eoi-definition}, we captured 156 scripted events across 10 campus zones and 10 day/night sessions with 22 student actors, resulting in 20 non-stop patrol videos around 30 minutes each, containing 833 people and 1,537 vehicles.
For video capture, we utilized a Frodo Zero ground robot~\cite{frodobots} equipped with a GoPro Hero 11~\cite{goprohero11}, recording 4K-resolution videos without audio during video taking. 
Each video contains GPS coordinates and timestamps. 
\Three{Our video production process is as follows.}

\Three{\textbf{Step 1. Preparation}
Before the main video production, we prepared four elements to prioritize ecological validity, ensuring footage appeared natural, realistic, and reusable for future studies.
Specifically, we focused on defining plausible patrol routes, creating naturalistic acting guidelines, conducting dry-run pilots, and securing public safety professional oversight.
First, robot patrol routes were designed to cover campus ``hot spots'' identified by the eighth author with public-safety domain expertise, remain fully wheel-accessible, and avoid spatial overlap.
Second, the acting scripts specified naturalistic behavior, such as maintaining a normal walking pace, avoiding exaggerated glances toward the robot, and performing each EoI at varying distances, locations, and occlusion levels to reflect real-world visual complexity.
We balanced the distribution of EoI types and ensured that each EoI was represented at least once in the dataset.
Third, we conducted dry-run recordings using two scripted routes and seven actors, reviewing two 30-minute pilot videos to tune robot speed, route length, and ensure actor behaviors appeared natural.
Finally, the eighth author confirmed these behaviors aligned with actual campus incidents, leading us to refine both patrol routes and acting scripts before final production. 
\begin{figure*}
  \centering
  \includegraphics[width=\linewidth]{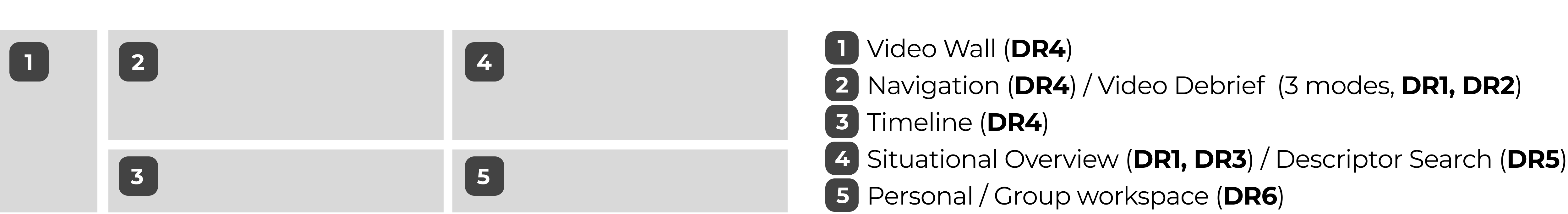}
  \caption{\system interface layout and corresponding design requirements.}
    \label{fig:UI}
\end{figure*}
\label{section:testbed}
We note that some abnormal events cannot be naturally captured, so we followed prior work~\cite{Lu2025CLIPSENetCS, souza2017phav} using actor-performed scenarios to ensure systematic coverage of rare and critical events.
}

\Three{\textbf{Step 2. Video Capture}
Before each capture, actors walked through their assigned locations and were briefed on where, when, and how to perform each EoI naturally within everyday campus settings.
To synchronize events with the robot’s continuously moving field of view, a coordinator followed the robot (out of frame) and provided silent temporal cues, ensuring that events unfolded as the robot arrived without actors breaking character or visibly ``waiting'' for the camera.
All recordings took place in an active, unannounced campus environment: while the \textit{foreground} events were scripted, the \textit{background} included uncontrolled pedestrian traffic, vehicles, lighting variation, and other ambient activities to preserve ecological validity.}
After collecting the videos, we annotated each recording with ground-truth labels, including EoI types, timestamps, and event durations.

\Three{To simulate a scenario in which ten videos are captured simultaneously, we designed spatially non-overlapping patrol routes, each assigned a spatiotemporal index (route ID and timestamp ID) corresponding to an individual video.
We collected ten patrol sessions using the same robot across these routes over ten days under comparable weather conditions and at the same time of day, thereby emulating simultaneous multi-robot coverage of the environment; all recordings were completed within one month to maintain consistent seasonal lighting and foliage.
For each route, we recorded a daytime and nighttime pair of videos.
To avoid unrealistic identity collisions, actors changed outfits across sessions, and scripts were designed to prevent impossible location jumps or the same individual appearing simultaneously in different videos.
Finally, three authors and four volunteers reviewed all videos and scripts to confirm the absence of cross-video identity conflicts and to ensure that appearances across days remained visually distinguishable.
}

\Three{\textbf{Step 3. Post Production: Anonymization and Distribution}}
All identifiable faces and license plates were blurred using ORB-HD Deface~\cite{ORB-HD_deface_2025} for facial anonymization and EgoBlur~\cite{raina2023egoblur}, which is based on Faster R-CNN~\cite{ren2016fasterrcnnrealtimeobject}, for license-plate anonymization.
This produces a research-safe foundation for evaluating timeline-based filtering, real-time alerting, descriptor-based search, and event detection. The dataset contains video only, with no audio, to avoid potential First Amendment concerns~\cite{freedom}.
\Three{A research-safe version will be shared publicly, including (a) anonymized video files (with all faces and license plates blurred) and (b) metadata describing EoI types, timestamps, and GPS information.}

\section{\system}
We built \system to faithfully embody the DRs identified in S1. 
Developing such a tool posed the two key challenges~\Three{raised consistently across S1 themes and echoed in prior work.}
First, providing scalable real-time insights from video investigations requires a strong backend capable of balancing speed and reliability~\Three{\cite{zhang2017live, isenkul2025energy}}.
\Three{In S1, professionals described the "grind" of manually reviewing massive, dispersed footage under time pressure (T2, T3). 
Unreliable automated results force additional 
verification work, compounding operator burnout~\cite{parasuraman1997humans, haque2024we}.}
Second, delivering these insights to public safety professionals demands a simple and trustworthy flow of information; the system must minimize false alarms yet surface all noteworthy cases so that no critical issue is overlooked~\Three{\cite{kumar2018rethinking, tariq2025alert}}.
\Three{In high-stakes scenarios, professionals stressed the need to understand why context-aware events are flagged (T1, T4), as false alarms, and opaque reasoning erode trust in safety-critical alerting systems~\cite{dixon2006automation, liao2020questioning, mastrianni2022pop}.}
Addressing these challenges required balancing two tensions: building an advanced backend generating all possible cases, while presenting them to avoid overwhelming users.
Our novelty lies in balancing the two demands: advancing video understanding models to be fast, accurate, and trustworthy, and structuring their output into clear, actionable insights through the front-end.
To this end, we introduce the backend pipeline leveraging \textit{multimodal LLM} prompt-engineered with the EoI taxonomy and professionals’ domain-specific analysis protocol, and a front-end that presents noteworthy events selectively in a \textit{structured format} of (1) event type, (2) explanation, (3) rationale, and (4) confidence level.
Together, these components enable public safety professionals to build situational awareness across both area and time.
Embedding the DRs into our system, we organized them into four recurring user workflows:
\textbf{F1. }Browsing and investigating detected EoIs for situational awareness (DR1–DR3),
\textbf{F2. }Reasoning across time and space (DR4),
\textbf{F3. }Locating objects of interest through targeted search (DR5), and
\textbf{F4. }Collaborating with team members (DR6).
\begin{figure*}
  \centering
\captionsetup{justification=raggedright,singlelinecheck=false}
  \includegraphics[width=1\textwidth]{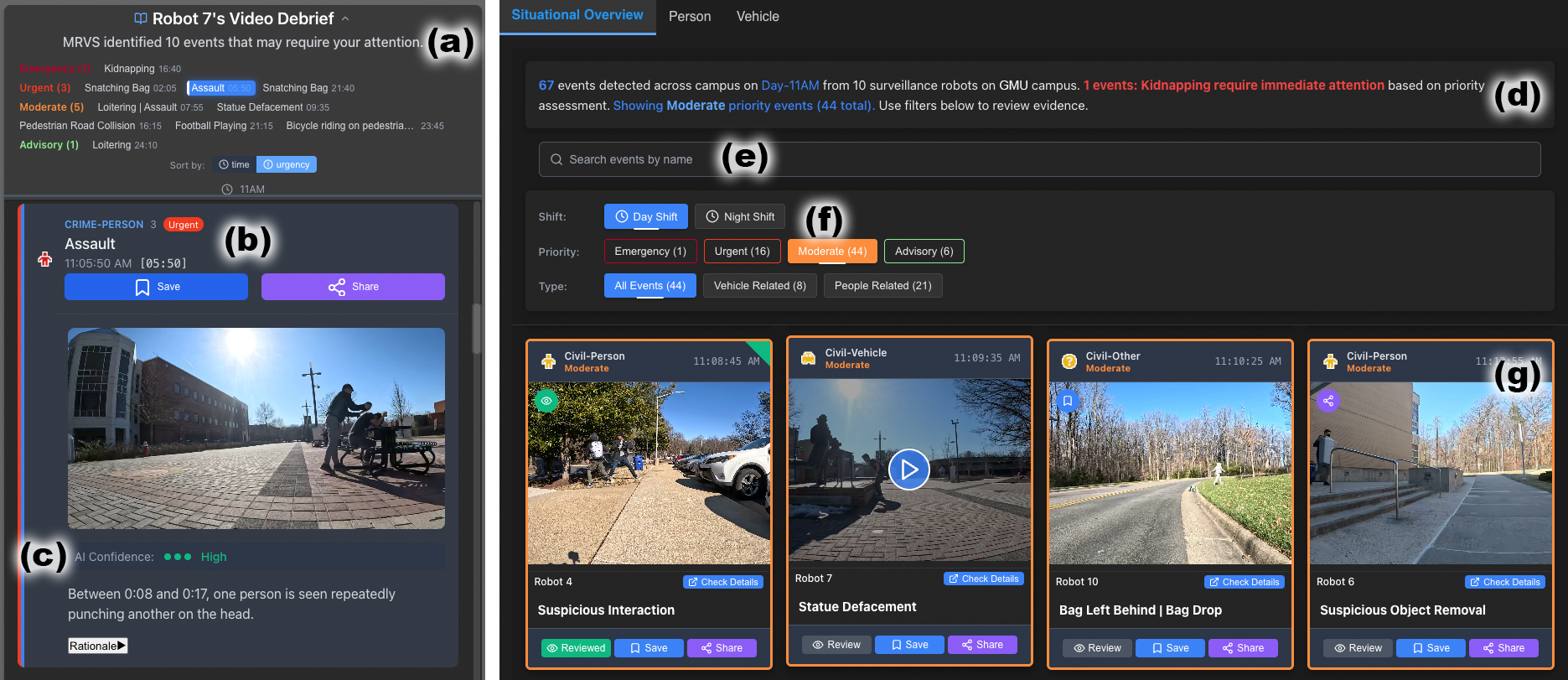}
  \caption{F1.Browsing and investigating detected EoIs for situational awareness. \Six{\textbf{Left}: Professionals begin with a robot-level video debrief \textit{(a)}, where detected events are grouped by priority and sorted by time/urgency. Selecting an event opens an inspectable card \textit{(b)} with triage actions (save/share), and a representative keyframe; with model confidence and rationale \textit{(c)}.
\textbf{Right}: The situational overview summarized events across robots \textit{(d)}, supports keyword search \textit{(e)}, and filtering by shift, priority, and event type \textit{(f)}. Filtered events are presented as cards for rapid scanning \textit{(g)} containing mark reviewed, save, or share items, quickly viewed event video segment, and clicking “Check Details” links a card to deeper inspection on the card.}}
  \label{fig:workflow1}
\end{figure*}
\begin{figure*}
    \centering
\captionsetup{justification=raggedright,singlelinecheck=false}
  \includegraphics[width=\linewidth]{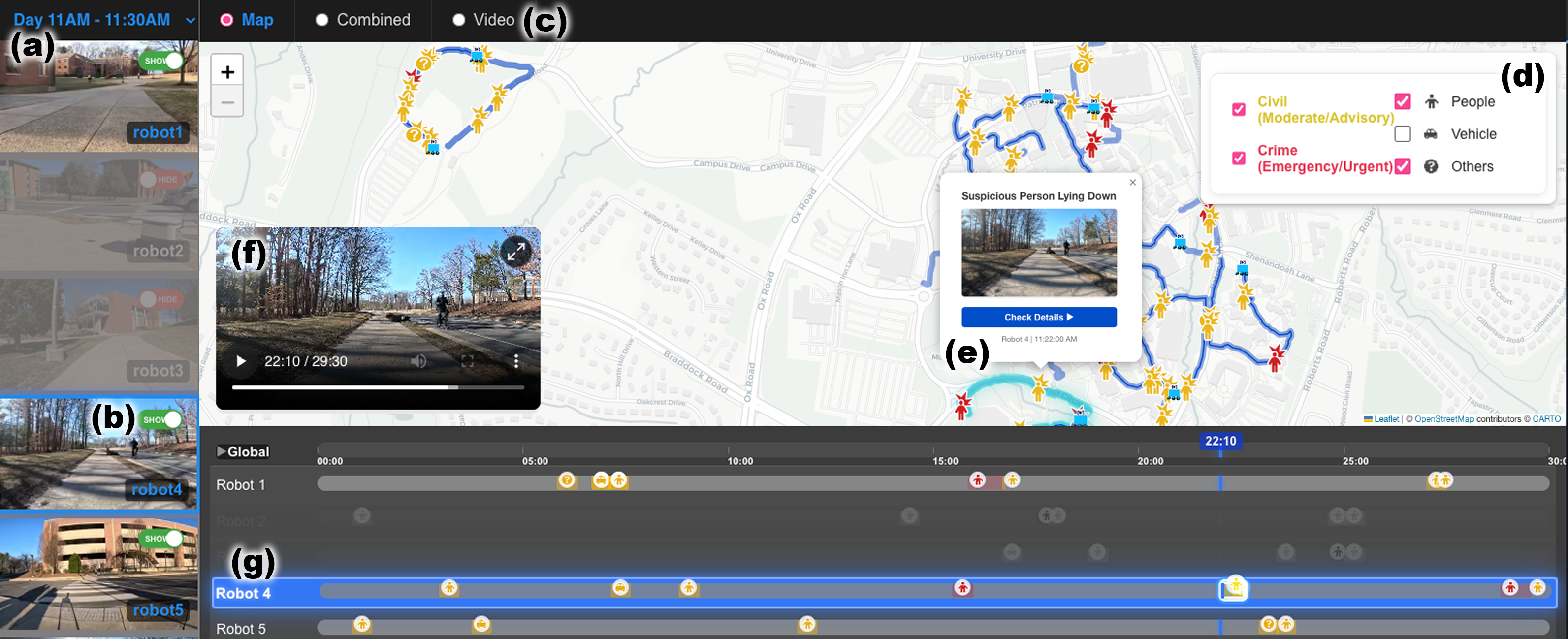}
\caption{
F2. Reasoning across time and space. Professionals adjust the day/night window time for videos \textit{(a)} and browse multiple robots via the video list, with toggles to show/hide videos linked to timeline and trajectory \textit{(b)}. Three layout options display map, video debrief, and video \textit{(c)}. The legend supports event type and entity filtering \textit{(d)}. Robot trajectories can be selected, with icons linking to keyframe popups and a ``Check Details'' for deeper inspection \textit{(e)}. Users can quickly preview corresponding video segments \textit{(f)}. A global timeline serves as a shared temporal reference \textit{(g)}, while per-robot timelines with synchronized playheads enable aligned cross-robot comparison and time-jump navigation for spatiotemporal reasoning.
}
    \label{fig:workflow2}
\end{figure*}
\subsection{Frontend: Interactive Sensemaking and Collaborative Investigation}
In this section, we explain how features in \system are designed, built, and embedded to support the four flows and design requirements. 
Our interface dashboard use five divisions (D1\textasciitilde D5) as shown in Fig.~\ref{fig:UI}.
In describing the four flows, we will explain how we utilized each division using the features.
\paragraph{F1. Browsing and investigating detected EoIs (DR1–DR3).}  
\system supports browsing through \textit{Video Debrief} (D2) and \textit{Situational Overview} (D4).  
\textbf{Video Debrief} (Fig.~\ref{fig:workflow1}, left) structures detected EoIs within a single video as chapter-like chunks, each with title, description, rationale, confidence score, and a representative frame. 
A prioritized summary and timeline navigation bar at the top allows quick access, while clicking the representative frame synchronizes all components to the event’s start time.
\textbf{Situational Overview} (Fig.~\ref{fig:workflow1}, right) aggregates EoIs across multiple videos, showing events as cards with robot ID, time, and priority.
Cards can be filtered, played, marked, or synchronized with the global timeline.  
These components reflect DR1-DR3, reducing manual video review and enhancing situational awareness by surfacing structured, contextual information from the back-end.
\begin{figure*}
  \centering
\captionsetup{justification=raggedright,singlelinecheck=false} 
  \includegraphics[width=1\textwidth]{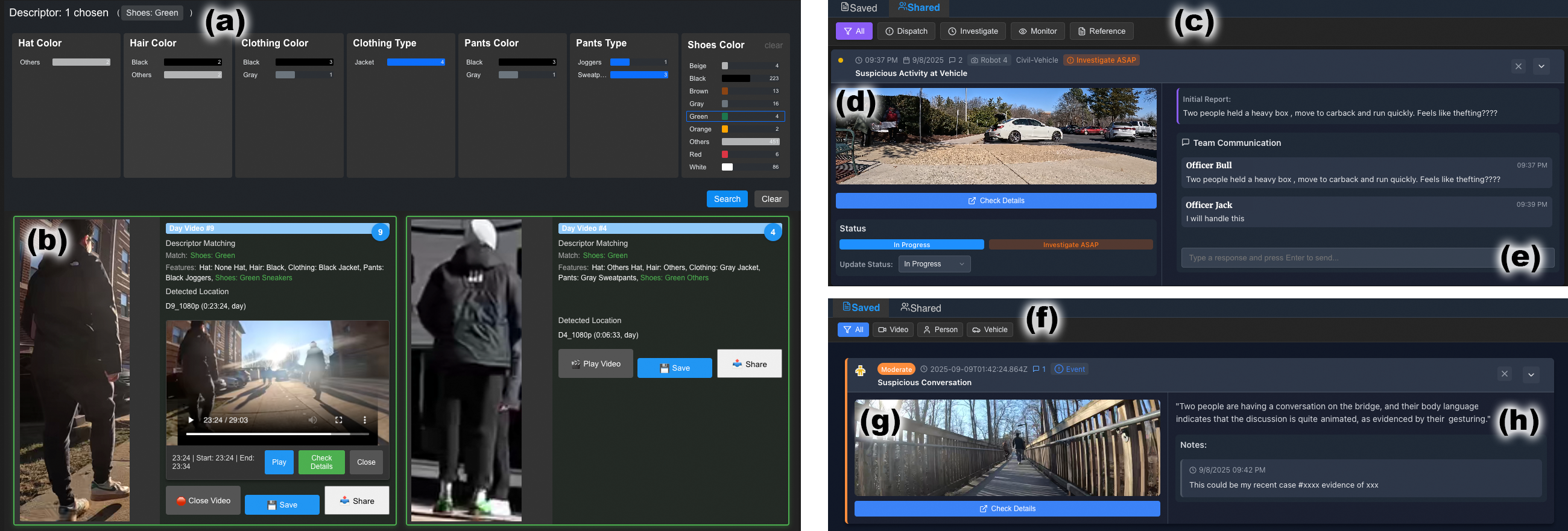}
\caption{\textbf{Left} F3. Locating objects of interest through targeted search. \Six{Professionals build queries using appearance descriptors \textit{(a)}; results appear as ranked cards with keyframes, matched attributes, and clip-level evidence for verification \textit{(b)}.} 
  \textbf{Right} F4. Collaborating with team members (DR6).
  \Six{Events can be shared to a team workspace \textit{(c)}, where items become trackable work cards with status/priority and evidence entry points \textit{(d)}. Team members coordinate via a shared communication thread \textit{(e)}. Saved events in personal workspace are organized by entity-level filter \textit{(f)}, displaying as detailed note cards with save time and evidence keyframes with entry points \textit{(g)} and model event descriptions with personal notes \textit{(h)}.}}
  \label{fig:workflow3}
\end{figure*}
\paragraph{F2. Reasoning across time and space (DR4).}  
To support spatio-temporal reasoning, \system links all feeds through a synchronized \textit{Video Timeline} (D3), \textit{Map Navigation} (D2), and the \textit{Video Wall}, together in Fig.~\ref{fig:workflow2}.
The \textbf{Video timeline} provides a global, color-coded view of EoI types and priorities, allowing professionals to scan activity, filter by category, and jump into synchronized playback. 
The map visualizes robot real-time updating trajectories with pinned EoIs locations; selecting a marker retrieves the corresponding video timestamp. 
\Three{Each EoI is represented by a simplified icon (person/vehicle/other) determined by its properties, positioned on the map using the robot’s GPS at the time of the event.}
All three components are interconnected.
D2 further supports three in-division layouts that differ in the location and size of the map navigator, video navigator, and video debrief for different analytical focuses. 
This reflects DR4 coupling and enables professionals to track how events emerge, propagate, and connect across both space and time.
\begin{figure*}
  \centering
  \includegraphics[width=\textwidth, clip, trim=00 18 00 00]{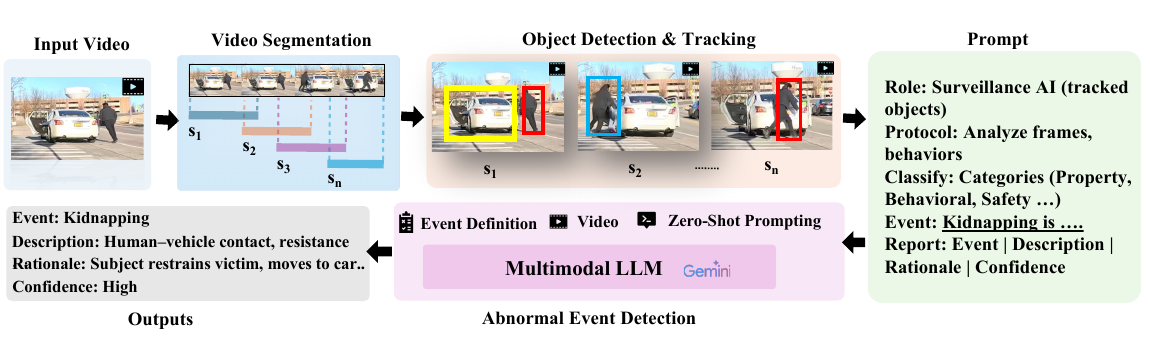}
  \caption{\system backend EoIs detection process, support front-end Video Debrief \& Situational Overview.}
  \label{fig:back-endsystem-architecture}
\end{figure*}
\paragraph{F3. Locating objects of interest through targeted search (DR5).}
\system enables targeted search through a \textbf{Descriptor Searcher} (D4) see Fig.~\ref{fig:workflow3}. Person and Vehicle appearance features, such as clothing, vehicle type, or color, are derived from back-end descriptors with visual shown. 
Search results are shown as cards with images, metadata, and match scores, each linked to the corresponding video clip. This reflects DR5 in filtering and comparison functions that help professionals narrow results and trace objects across multiple videos, reducing time spent on manual scanning.
\paragraph{F4. Collaborating with team members (DR6).}
Collaboration is supported through a \textbf{Group Workspace} (D5) see Fig.~\ref{fig:workflow3}, where professionals can save, annotate, and share EoI cards/chapters or search results.
Each entry keeps its contextual details: event type, time, and robot ID, so that teams can review evidence consistently. 
Professionals can leave notes, validate detections, and organize cases for follow-up, enabling shared situational awareness and coordinated decision-making for DR6.
\subsection{Back-end: MLLM-based Video Sensemaking and Descriptor Searching}
The back-end of \system converts testbed multi-robot footage into structured, reviewable information about EoIs and object visual attributes, and outperformed the state-of-the-art traditional computer vision anomaly detection model, HolmesVAD~\cite{zhang2024holmesvadunbiasedexplainablevideo} and LLM Gemini~\cite{geminiteam2024geminifamilyhighlycapable}. 
By combining multi-object detection and tracking with multimodal LLM reasoning, our back-end model could detect and classify EoIs, describe objects, generate confidence-scored event types with descriptions and rationales, extract key preview thumbnails, and generate stable appearance-based descriptors such as clothing or vehicle type, providing the foundation for front-end video debrief, situational awareness, and descriptor-based search.

\subsubsection{MLLM-based Video Sensemaking (DR1-DR3)}
To help professionals quickly grasp long videos, 
the \system back-end employs a \quotes{surveillance persona} LLM incorporating the EoIs taxonomy from S1 and follows domain-specific analysis protocols as an \quotes{expert analyst}. 
This persona reviews videos in short clips, reasons about key objects and activities, and generates structured event cards with event type, description, rationale, confidence level, and a representative frame. 
These outputs reduce professionals’ cognitive load and remove manual browsing. 
With the confidence level and AI-generated rationale, professionals can also assess how the AI reached its conclusions and evaluate the reliability of its evidence.

\Three{Specifically, our approach builds upon the capabilities of Gemini 2.0 Flash \cite{geminiteam2024geminifamilyhighlycapable} as a baseline based on its leading performance on established video understanding benchmarks \cite{wu2024longvideobenchbenchmarklongcontextinterleaved, wang2025lvbenchextremelongvideo, salehi2024actionatlasvideoqabenchmarkdomainspecialized} and enhances it with multi-modal reasoning and spatial-temporal localization. The model was configured with a temperature of 1.0 to balance response creativity with output consistency and reliability.} We guide attention to objects of interest with point prompting \cite{dai2024samaugpointpromptaugmentation, dai2024curriculumpointpromptingweaklysupervised}. Specifically, we preprocess the video streams using the BoxMOT \cite{Brostrom_BoxMOT_pluggable_SOTA} framework, which integrates real-time multi-object tracking with YOLO \cite{wang2024yolov10realtimeendtoendobject} based detectors. Each frame is enriched with bounding boxes $\mathcal{B} = \{b_i = (x_{1,i}, y_{1,i}, x_{2,i}, y_{2,i})\}_{i=1}^{N}$ and class labels $\mathcal{C} = \{c_i\}_{i=1}^{N}$ for detected entities.  These annotated frames form an object-aware video stream $\mathcal{V}^*$, which is then processed by a multimodal large language model (MLLM) using both \textbf{role prompting} and \textbf{temporal segmentation}. Inspired by \cite{kong2024better}, we introduce a system-level persona prompt, defining the model as a specialized surveillance agent (VSAI-9000) with domain-specific analysis protocols. This role-based formulation enhances zero-shot reasoning and structured event classification.
\Three{The input video is temporally segmented into overlapping intervals $\mathcal{S} = \{s_j\}_{j=1}^{n}$ (e.g., 10 seconds each), where consecutive segments share partial temporal overlap to ensure continuous coverage of key activities and smooth transition detection.} \Three{The 10 second interval duration was determined through pilot testing with 10 videos, balancing sufficient temporal context for meaningful event identification against computational efficiency and the model's context window limitations.} For each interval, the pipeline (Fig.~\ref{fig:back-endsystem-architecture}) returns the following: the event type \( e \in \mathcal{E} \), a natural language description \( \mathcal{D}(e) \), an explanatory rationale \( \mathcal{R}(e) \),  \Three{a confidence level \( l(e) \in \{\textit{high}, \textit{medium}, \textit{low}\} \)}, and representative frame(s) \( \mathcal{K}(e) \) selected via keyframe extraction. A keyframe is selected as the frame with the highest count of distinct objects. This frame serves as the clearest snapshot of the interval.   
\Two{The model is instructed to state its confidence as one of three discrete levels, \textit{high}, \textit{medium}, or \textit{low}, based on how certain it is about the inferred event type and description. This confidence level is passed unchanged to the front-end (e.g.~shown as ``High'' in Fig.~\ref{fig:UI}) and used as an ordinal indicator of reliability.
This design follows recent work showing that large language models can express reasonably calibrated uncertainty in natural language without exposing logits or explicit probability scores \cite{lin2022teaching,kadavath2022languagemodels}.}
\subsubsection{Descriptor-based Search (DR5)}  
To support targeted investigation, the system back-end provides a \quotes{descriptor-based search} function allowing professionals to locate specific people or vehicles from partial descriptions. It extracts stable visual attributes, such as clothing color or vehicle type and color, paired with cropped object images. This enables professionals to quickly filter and verify candidates without relying on memory or manual review.

Objects are extracted from videos using BoxMOT \cite{Brostrom_BoxMOT_pluggable_SOTA}, combining YOLO-based object detectors \cite{redmon2016you} with DeepOCSort \cite{maggiolino2023deep} for multi-object tracking. LLAVA 1.6 \cite{liu2024llavanext} verifies detection semantic correctness through binary questions, discarding negatives. For cars and persons, descriptor-based attributes are extracted through structured queries, with cropped first-appearance images stored for validation. Predicted attributes are re-evaluated on saved crops, and rejected outputs are placed in the \quotes{other} category.

\Three{Descriptor-based search matches the selected car or person against all objects sharing similar descriptor values. We extract descriptors with LLAVA~1.6~\cite{liu2024llavanext}, which chooses from predefined options or answers yes/no. For non-binary descriptors, we include \quotes{other} and \quotes{unclear} so low-confidence cases avoid overly specific labels. We then run a second yes/no round (converting option questions to binary) to filter out low-confidence answers. This removes low-confidence answers and improves robustness by reducing the impact of weak predictions on search accuracy.}

To address missed tracking, image similarity checks with a contrastive model \cite{tschannen2025siglip} compare same-class objects sharing identical descriptors (shirt and pants color for people; body color for vehicles). Trajectories are merged when similarity surpasses a high threshold (0.95), with final descriptor assignments resolved through majority voting. The system preserves the detection crop exhibiting the maximum bounding box area per trajectory.

\section{Study 2: Summative Study}
In this summative study (S2), we conducted two evaluations assessing \system's practical utility for back-end technical performance and public safety professionals' video sensemaking tasks. 
In algorithmic performance evaluation, we compared two state-of-the-art anomaly detection models as baselines against our improved LLM-enhanced module.
We conducted an expert review with 9 public safety professionals at their respective offices to evaluate how the system’s frontend supports real-world investigative workflows.
\begin{table*}[!t]
\centering
\footnotesize
\begin{tabular}{lcccccccc}
\toprule
\textbf{Period} & \textbf{\# Videos} & \textbf{Duration (min)} & \textbf{Normal} & \textbf{Abnormal} & \textbf{Method} & \textbf{Precision} & \textbf{Recall} & \textbf{F1} \\
\midrule
\multirow{3}{*}{Day} 
  & \multirow{3}{*}{10} & \multirow{3}{*}{250} & \multirow{3}{*}{400} & \multirow{3}{*}{218} 
  & HolmesVAD~\cite{zhang2024holmesvadunbiasedexplainablevideo} & 0.223 & 0.008 & 0.016 \\
  &                    &                       &                       &                       & Gemini 2.0 \cite{geminiteam2024geminifamilyhighlycapable}      & \textbf{0.769} & 0.351 & 0.469 \\
  &                    &                       &                       &                       & Ours             & 0.505 & \textbf{0.534} & \textbf{0.497} \\
\midrule
\multirow{3}{*}{Night} 
  & \multirow{3}{*}{10} & \multirow{3}{*}{220} & \multirow{3}{*}{404} & \multirow{3}{*}{161} 
  & HolmesVAD~\cite{zhang2024holmesvadunbiasedexplainablevideo} & 0.074 & 0.012 & 0.021 \\
  &                    &                       &                       &                       & Gemini 2.0 \cite{geminiteam2024geminifamilyhighlycapable}        & \textbf{0.623} & 0.368 & 0.438 \\
  &                    &                       &                       &                       & Ours             & 0.440 & \textbf{0.792} & \textbf{0.540} \\
\midrule
\multirow{3}{*}{Overall}
  & \multirow{3}{*}{20} & \multirow{3}{*}{470} & \multirow{3}{*}{804} & \multirow{3}{*}{379} 
  & HolmesVAD~\cite{zhang2024holmesvadunbiasedexplainablevideo} & 0.015 & 0.011 & 0.018 \\
  &                    &                       &                       &                       & Gemini 2.0 \cite{geminiteam2024geminifamilyhighlycapable}        & \textbf{0.696} & 0.359 & 0.453 \\
  &                    &                       &                       &                       & Ours             & 0.473 & \textbf{0.663} & \textbf{0.519} \\
\bottomrule
\end{tabular}
\caption{Detection performance on 20 videos (470 min). Best scores per metric (per period) are in bold.}
\label{tab:results}
\end{table*}
\subsection{\Four{Algorithm Evaluation}}
\subsubsection{Method}
To emulate the sliding-window logic of an online detector, we divide every video into \SI{30}{\second} segments with a \SI{5}{\second} overlap; therefore, a new segment starts every \SI{25}{\second}.
We denote each segment by $s_i$ where $i = 1,\dots,N$ and $N$ is the total number of segments across all videos. 
For each segment $s_i$ the proposed model generates a text prompt that is matched against a predefined \emph{incident taxonomy} comprising eight high-level categories (\emph{Vehicle \& Mobility}, \emph{Public-Order Disturbance}, \emph{High-Risk Threat}, \emph{Suspicious Behaviour}, etc.).
Each category contains several fine-grained events, e.g. \emph{Hit-and-Run}, \emph{Loitering}, or \emph{Shooting}. The pipeline outputs at most one incident label per segment.
Three trained annotators independently labeled every segment. A segment was marked \emph{abnormal} only if at least two annotators agreed that some event from the taxonomy was visible; otherwise it was marked \emph{normal}.
Comparing the model prediction with the ground truth yields four mutually exclusive outcomes per segment: True Positive (TP), False Positive (FP), True Negative (TN), and False Negative (FN).

The four counts above form the confusion matrix. \Four{Because abnormal events are relatively rare compared to normal segments (Table~\ref{tab:results}), overall accuracy is misleading.
For example, a system that always predicts ``normal'' would achieve high accuracy but zero utility for anomaly detection. 
We therefore report three standard evaluation metrics used in information retrieval, classification, and video anomaly detection under class imbalance \cite{manning2008introduction,davis2006relationship,Sultani_2018_CVPR}:
} 
\begin{itemize}
    \item Precision (P), \Four{defined as $\mathrm{TP}/(\mathrm{TP}+\mathrm{FP})$, measures the proportion of segments predicted as abnormal that are truly abnormal. High precision reduces false alarms and unnecessary verification effort.}
    \item Recall (R), \Four{defined as $\mathrm{TP}/(\mathrm{TP}+\mathrm{FN})$, measures the proportion of truly abnormal segments that are successfully detected. High recall is critical in safety-critical surveillance, where missed incidents can have serious consequences.}
    \item F1 Score, \Four{defined as the harmonic mean of P and R, $2PR/(P+R)$, summarizes the trade-off between precision and recall into a single number and penalizes systems that perform well on only one of the two \cite{manning2008introduction,davis2006relationship}. In video anomaly detection, where positives are rare and both false alarms and missed detections matter, F1 is widely used as a primary comparison metric \cite{Sultani_2018_CVPR}.}
\end{itemize}

To benchmark our method, we compare against two baselines: (1) HolmesVAD~\cite{zhang2024holmesvadunbiasedexplainablevideo}, a recent state-of-the-art model specifically designed for explainable video anomaly detection, and (2) Gemini 2.0~\cite{geminiteam2024geminifamilyhighlycapable}, a powerful multimodal large language model, which we adapt as a strong baseline for anomaly detection by leveraging its multimodal reasoning capabilities. This setup enables a meaningful comparison between specialized video models and the emerging class of general-purpose LLMs.

\subsubsection{Results}
Table~\ref{tab:results} summarizes the detection performance across 20 videos (470 minutes), broken down by time of day (Day and Night). For each method, we report precision, recall, and F1 score, with the best value per metric highlighted in bold. Our method’s ability to detect nearly twice as many anomalies results in substantially better overall detection performance and practical utility. It consistently outperforms both HolmesVAD, a state-of-the-art video anomaly detection model, and Gemini 2.0, a strong multimodal LLM baseline we adapted for this task. During the day, our method achieves the highest recall (0.534) and F1 score (0.497), successfully balancing the precision-recall tradeoff—unlike Gemini 2.0, which favors precision (0.769), suffers from much lower recall (0.351), ultimately missing a significant number of anomalies. At night, the superiority of our approach becomes even more evident, achieving the best precision (0.440), recall (0.792), and F1 score (0.540). In contrast, both baselines deteriorate sharply, especially HolmesVAD, which drops to a recall of just 0.012. Overall, our method achieves the highest F1 score (0.519) and recall (0.663), confirming its robustness across diverse lighting conditions. These results validate our design choices and demonstrate that a task-specialized approach significantly outperforms both specialized and general-purpose multimodal baselines in real-world anomaly detection scenarios.

\Four{The lower precision of our method compared to Gemini 2.0 (e.g.~0.505 vs.\ 0.769 during the day and 0.440 vs.\ 0.623 at night) reflects an intentional design trade-off.
We configured MRVS to more liberal in proposing candidate anomalies so as to minimize false negatives in our safety-critical setting. Essentially, extra false positives increase screening workload for public safety professionals where missed incidents correspond to unobserved threats. 
In practice, MRVS is intended to be used as a decision-support tool in which flagged segments are reviewed and validated by officers. In this setting, the substantial gains in recall and F1 are more important than maximizing precision alone.}
\subsection{\Four{Expert Review}} 
The goal of expert review was to address our research questions regarding professionals’ perceptions, use experiences, and reflections on the adoption of \system in their investigative workflows.
Specifically, this study aimed to explore:
\begin{itemize}
    \item \textbf{RQ1}: How might the adoption of the MRVS system transform current police agency practices, including expected benefits, risks, and its influence on team collaboration?
    \item \textbf{RQ2}: How do professionals perceive the functionality of each system component, including AI-generated debriefs and descriptor-based object tracking?
    \item \textbf{RQ3}: What are the considerations, challenges, and design opportunities for integrating MRVS-like systems into future public safety workflows?
\end{itemize}

\subsubsection{Participants and Recruitment.}
We recruited nine experienced public safety professionals (P1–P9) from five local police departments, including patrol officers, detectives, captains, and real-time crime center officers.
All participants had at least 5 years of field experience (see Table~\ref{tab:participant-s2}) and were familiar with video investigation in their daily work, but had limited exposure to AI-augmented video analysis systems.
Participants were recruited through institutional partnerships and volunteered to participate without compensation.
\begin{table*}[t]
\centering
\footnotesize
\begin{tabular}{lllrp{6cm}}
\toprule
\textbf{PID} & \textbf{Rank} & \textbf{Agency} & \textbf{Years} & \textbf{Primary Focus Areas} \\
\midrule
P1 & Detective & George Mason University Police Department & 7 & Post-incident investigation \\
P2 & Detective & Manassas City Police Department & 12 & Post-incident investigation \\
P3 & Detective & Manassas City Police Department & 8 & Post-incident investigation, drones \\
P4 & Sergeant & Virginia State Police Department & 16 & Real-time crime, post-incident investigation, drones \\
P5 & Detective & City of Fairfax Police Department & 5 & Post-incident investigation, dispatcher, patrol \\
P6 & Detective & City of Fairfax Police Department & 25 & Post-incident investigation \\
P7 & Captain & City of Fairfax Police Department & 18 & Patrol, criminal investigation, community service \\
P8 & Captain & Fairfax County Police Department & 16 & Real-time crime, community service \\
P9 & Sergeant & George Mason University Police Department & 13 & Patrol, real-time crime \\
\bottomrule
\end{tabular}
\caption{Study 2 Participant Profiles.}
\label{tab:participant-s2}
\end{table*}
\vspace{-1em}
\subsubsection{Procedure.}
We designed a structured evaluation procedure combining task-based expert use of \system with post-task interviews to assess its utility in realistic policing scenarios.
Sessions were conducted in person with nine public safety professionals at their offices, using our monitor for consistency. 
After a brief introduction and demonstration of \system’s core components: multi-robot video playback with timelines and maps, LLM-powered debrief and situational overview, clickable color-coded icons, descriptor-based search, and a collaborative workspace—participants were invited to freely explore the system to gain familiarity. 
They were then asked to complete three investigative tasks simulating realistic police scenarios, followed by \textbf{semi-structured interviews} examining how \system compared with current workflows (RQ1), perceptions of specific system components (RQ2), and broader views on MRVS's role in future policing (RQ3).
Sessions lasted between \textbf{75 and 120 minutes}.
\Four{Given the open-ended, scenario-based task design, participants used \system's features in varying combinations and sequences. 
Consequently, we report findings as a holistic system evaluation rather than a comparative analysis of features across tasks.
}
\begin{itemize}
    \item \textbf{Task 1}: Urgent Incident Response and Verification. 
    Locate and verify video evidence of an ongoing fight reported on a campus bridge, saving key findings with notes for team coordination. 
    \item \textbf{Task 2}: Routine Patrol Review and Event Validation. 
    Review AI-detected events from a single time shift, validate significant events versus false alarms, and share verified cases for further action (15-minute session). 
    \item \textbf{Task 3}: Descriptor-based Search and Suspect/Vehicle Identification.
    Use witness descriptions to search, locate, and verify suspects or vehicles, sharing matches to support investigations.  
\end{itemize}
\subsubsection{\Four{Expert Interview Analysis}}
\Four{We analyzed the expert interviews using a data-driven thematic analysis approach~\cite{muller2014curiosity, saldana2015coding}. 
All sessions were audio recorded, transcribed independently open-coded by two trained authors using the Glaser method, with short memos capturing contextual details~\cite{glasser1992basics}.
To validate coding quality, we assessed inter-rater reliability on the pre-reconciliation codes, achieving substantial agreement (Cohen’s $\kappa$ = 0.78). 
They then compared codes and iteratively clustered them into higher-level categories via constant comparison~\cite{saldana2015coding}, resolving ambiguities by revisiting transcripts and, when needed, consulting the corresponding author.
Together, the two coders and the corresponding author diagrammed relationships among categories and refined them into themes organized around our three research questions (RQ1-RQ3).}

\subsubsection{Front-end evaluation results}
\paragraph{\textbf{RQ1:~\system Overview}~Empowers Police Workflows and Strengthens Team Collaboration}
Participants overwhelmingly described \system as essential support for the cognitively demanding work of video sensemaking, both in real-time dispatch and post-incident investigations. 
Professionals emphasized that analyzing fragmented, multi-source footage, whether real-time or post-incident, was among the most exhausting and resource-intensive aspects of their work, often requiring hours of scanning and attention to subtle cues~\Four{(N=7/9)}.
\system was seen as a \emph{force multiplier} that eased these burdens, helping them prioritize attention, surface relevant segments, and focus on verification and decision-making amid staff shortages~\Four{(N=7/9)}. 
Professionals noted that by minimizing time spent on exhaustive searches, the system freed them to focus on higher-value interpretive tasks critical to effective investigations.
Beyond supporting individual sensemaking, \system also addressed longstanding challenges in coordinating video investigations across shifts, teams, and timeframes. 
Professionals explained that cases often unfolded over extended periods, requiring asynchronous handovers and collaborative sensemaking~\Four{(N=3/9)}. 
Existing tools offered little support for this, leading to duplicated efforts, missed leads, or inconsistent interpretations~\Four{(N=5/9)}. 
The system’s collaborative workspace, including tagging, annotations, action items, and shared status updates, enabled teams to continue work seamlessly across shifts and off-days, supporting situational continuity and maintaining accountability without relying solely on informal or verbal handoffs.

However, these perceived efficiencies came with clear expectations for procedural integrity and accountability~\Four{(N=3/9)}.
Participants emphasized that AI-driven features, such as suspect identification and abnormal events detection, were valuable aids for narrowing focus during investigations but could not displace human judgment~\Four{(N=5/9)}. 
Professionals further advocated for safeguards to ensure legitimate use~\Four{(N=4/9)}, such as requiring case numbers before searches (P1), enforcing strict access controls (P2), and conducting regular audits (P4, P6), emphasizing aligning the system with broader legal and institutional protocols.
These accounts highlight how professionals negotiated the system’s role as a support for existing video sensemaking practices, embracing its ability to streamline workflows and strengthen team coordination, while insisting such systems operate within the procedural, legal, and professional boundaries central to public safety work.
\Four{Participants described \system as \quotes{easy to learn (P1, P3-P8)} and highly intuitive, all professionals completed the three evaluation tasks without assistance. They highlighted the natural interactive flow \quotes{I don't need to go back to the video, just click what I'm investigating on the timeline, map, or search, and the video jumps there (P4)} and fit to their workflow \quotes{Everything I need is right there, easy to find the alert, check it, and share through right near clicks (P7).}
Meanwhile, professionals (N=5/9) recommend splitting the current interface into two coordinated views to better support smaller monitors.}
\paragraph{\textbf{RQ2:~LLM-based Video Debriefs} Support Rapid Evidence Review while Reinforcing Public Safety Professionals Control and Trust}
Participants widely viewed AI-generated video debriefs as valuable accelerators for routine evidence review, particularly under high-pressure, resource-constrained conditions. Professionals appreciated features such as direct video jumps~\Four{(N=5/9)} and AI-curated summaries~\Four{(N=7/9)} for reducing cognitive load and enabling faster triage, while emphasizing these tools must operate within user-controlled workflows to maintain trust and accountability~\Four{(N=5/9)}. 
As P6 noted, \quotes{Even when it's wrong, it saves time because I can jump right to the clip and watch the footage.}
AI-generated confidence scores were especially valued for guiding attention and helping professionals calibrate their review effort: \quotes{If it's low confidence, I know I need to double-check everything carefully; if it's high, I can rely more on the detection and just validate (P1)}. 
Participants also welcomed transparent AI explanations embedded within the debriefs to help them assess system reasoning, even if imperfect~\Four{(N=6/9)}.
However, professionals cautioned that AI should remain a support tool, never replacing human judgment: \quotes{AI can't testify in court; we still have to make the final call (P3)}. 
Several participants raised legal concerns~\Four{(N=4/9)} about over-reliance on AI-curated evidence potentially introducing risks in court proceedings (\textit{P3}).

Participants framed MRVS as a user-controlled system where AI accelerates routine tasks and enhances situational awareness, while ensuring critical judgments, prioritization, and accountability firmly remain with the user. 
Some professionals further recommended the system allow setting custom alerts for high-risk incidents like assaults, with human oversight prioritized even in low-confidence detections (\textit{P4, P7}).
\Four{Although \system sometimes surfaced incorrect or missed events, all participants felt it substantially reduced workload versus exhaustive manual review.
The metadata-rich cards (keyframes, icons, event types, confidence rationales; Fig.~\ref{fig:workflow1}, left) let them dismiss many false alarms ``at a glance''~(N=6/9): \quotes{I realized this was wrong just look the image (P3)}.
For missed events, officers valued that the detected events, together with the synchronized map–timeline view, provided a spatial–temporal scaffold narrowed manual checks to specific time windows or locations \quotes{even when it misses one, I know exactly where to start(P6).}
}
\Four{
Beyond reducing workload, participants noted recurring failure modes in the underlying detectors: many missed events were \textit{small, brief, poorly lit, or near frame edges}, especially from ground-robot viewpoints where important behaviors occurred in the periphery rather than center (\textit{P3, P6, P9}).
These patterns led officers to treat MRVS alerts as useful starting points rather than exhaustive coverage, reinforcing the need for continued human review of subtle or peripheral incidents.}
\vspace{-2em}
\paragraph{\textbf{RQ2:~Descriptor-Based Search} is Valuable for Targeted Investigation}
Public safety professionals described the descriptor-based search feature as a powerful capability for narrowing suspects and focusing attention on relevant entities.
Participants valued AI’s support in filtering vehicles and persons, recognizing that even imperfect narrowing saved significant effort~\Four{(N=5/9)}: \quotes{Even if it just helps me rule out a few wrong ones, that's a huge help (P1)}. 
Professionals accepted that AI is better at broad categories (e.g., car type, color) than fine-grained details (e.g., brand)~\Four{(N=4/9)}, and recommended providing flexible filtering mechanisms, such as the ability to exclude certain features rather than only include them (P3, P5).
\Four{Beyond people and vehicles, participants emphasized the need to search for and track specific objects such as backpacks, bags, or items left behind across hours of footage~\Four{(N=2/9)}.}

Participants also stressed the need for progressive filtering workflows, starting from broad descriptors—like light vs. dark vehicles—before narrowing to specific details, reflecting the often vague information provided in real-world situations (\textit{P4}). 
However, they consistently emphasized that human interpretation remains essential for complex identifications and ambiguous cases~\Four{(N=7/9)}: \quotes{You still have to eyeball it yourself when things look alike. The system can’t make those calls for you (P3)}.
To support situational awareness, professionals recommended integrating search results directly into the map, allowing them to visualize suspects or vehicles over space and time: \quotes{If every time I do a search, the matches show up on the map, I can start building the suspect's trail right away (P7)}. 
While professionals acknowledged trade-offs between AI accuracy and workload reduction, they stressed the system must avoid excessive wrong detection~\Four{(N=3/9)}, with low-confidence results always requiring manual verification to preserve procedural rigor (\textit{P4}).
\paragraph{\textbf{RQ2:~Situational Overview} Enhances Situational Awareness and Professionals Safety}
Participants consistently described the Situational Overview as a valuable feature for enhancing situational awareness and officer safety by enabling rapid comprehension of complex scenes. Professionals appreciated the card-style key images and synchronized timeline and map functions, which allowed them to decide when to skim, when to validate, and when to investigate in depth~\Four{(N=5/9)}. 
This flexibility supported their preference for workflows where they \quotes{quickly scan and choose what to focus on} rather than following linear review processes.

Participants valued the system's filtering and marking capabilities~\Four{(N=4/9)}, empowering them to adjust the interface to investigative priorities and felt \system provided \quotes{a lot of options to sort and manage} while keeping them in control (\textit{P9}). 
Professionals emphasized that AI-assisted situational awareness should remain transparent and user-led, ensuring they retained discretion over how to explore, interpret, and act on information.
Professionals suggested future enhancements, such as question-based video querying, to streamline information retrieval without sacrificing context or control~\Four{(N=2/9)}. 
Overall, professionals framed the Situational Overview as a user-controlled, AI-assisted tool reducing cognitive load while supporting proactive and reactive policing tasks.
\paragraph{\textbf{RQ2: Collaboration Workspace} Supporting Flexible Police Practices}
Participants described current collaboration practices as fragmented and inefficient, relying on calls, emails, and paper trails that often led to information loss and uncertainty. 
They saw \system as a valuable opportunity to centralize and streamline coordination through a shared workspace, particularly praising the envisioned status update function as critical for cross-role and cross-agency collaboration (\textit{P1–P8}). 
Professionals highlighted the benefits of packaging case materials for easy sharing (\textit{P3}), but stressed the need for police-centered collaboration workflows~\Four{(N=5/9)}, including tiered permissions(\textit{P2, P5}), secure sharing with external partners(\textit{P1, P4}), and safeguards to ensure accountability(\textit{P3}).

They recommended features such as video sharing links (\textit{P1, P5}), downloadable reports (\textit{P2}), and case folders integrating all related events and clips (\textit{P4, P5}), ensuring supervisors could monitor progress without disrupting investigators. 
Importantly, professionals cautioned that AI-generated events should first undergo human review to avoid overwhelming frontline professionals (\textit{P9}). 
They emphasized balancing awareness with workload by tailoring notification settings by role~\Four{(N=4/9)}. 
Across these suggestions, participants underscored \system must support controlled, case-centered collaboration while minimizing friction and protecting rigor.
\paragraph{\textbf{RQ2: Seamless Spatial and Temporal Information}}  
Participants emphasized the importance of tightly integrated spatial and temporal information to support rapid situational awareness and investigation workflows (P1-P4, P8). 
While they appreciated \system's current map and timeline features, they identified key areas for improvement to better align with policing practices and decision-making under time pressure. 
They emphasized that visual icons and color-coding provide a clear overview of location and time without even seeing any text~\Four{(N=6/9)}. 

Participants highlighted the need for clearer geographical context, recommending that robot identifiers use location-based names rather than numerical codes to help locate video sources~\Four{(N=4/9)}.
They also proposed enhancing map interactivity by enabling direct selection of areas of interest, clickable patrol routes for instant timeline synchronization (P1-P3, P6-P8), and automatic highlighting of related events and movements when selecting a suspect or vehicle in descriptor-based searches~\Four{(N=3/9)}.
Overall, participants urged that future designs prioritize intuitive, police-centered interactions that seamlessly connect spatial and temporal data for efficient incident response.
\paragraph{{\textbf{RQ3: Envisioned Personalized MRVS} Supporting Flexible Police Practices}}
Professionals consistently envisioned MRVS not as a one-size-fits-all tool, but as a modular, adaptable workspace that aligns with their situational needs and investigative styles.
They emphasized the system should mirror their workflows, enabling adjustments to robot labels, event priorities, alert thresholds, and interface layouts~\Four{(N=6/9)}. 
For example, patrol robots should be named by location (e.g., “North Plaza”) to support situational awareness, and EoIs should be user-adjustable~\Four{plug and play module each agency could tailor to its policies, staffing levels, and local context (e.g., time of day, crowd density)}~\Four{(N=2/9)}.
Participants also highlighted that collaboration features should reduce reporting effort between colleagues and supervisors by making it easy to see who is working on which case and how far each has progressed~\Four{(N=3/9)}.

Professionals emphasized retaining agency and adaptive control over AI-generated alerts, preferring user-defined filters or role-based templates to avoid overload, especially in frontline tasks.
Investigators, supervisors, and real-time crime center officers were seen as needing different default views, filters, and notification templates for which alerts appear, how they are grouped, and how much detail is shown~\Four{(N=2/9)}.
They envisioned a minimalist, customizable interface where components like timeline or descriptor search could be rearranged, only expanded as needed. 
Many preferred distributing the interface across multiple screens to separate functions (e.g., real-time alerts, communications, post-incident review), aligning with existing dual-monitor setups~\Four{(N=6/9)}.
This vision reflects the expectation \system should adapt to professionals’ practices, supporting flexibility, transparency, and discretion, while preserving autonomy and accountability in public safety.
\Four{\paragraph{\textbf{RQ3: Envisioned Deployment and Human Robot Interaction (HRI)} Preventive, Hot-Spot-Oriented, and Specialized Use Cases}
Professionals described \system-like systems as primarily preventive resources, highlighting value in detecting emerging risks and discouraging undesirable behavior rather than only reconstructing incidents~\Four{(N=3/9)}.
They pointed to use cases such as flagging deteriorating infrastructure and repeatedly frequented locations where people \quotes{hang around for no good reason}, noting that visible patrol robots could help disperse groups while complementing traditional patrol work.\
Participants emphasized that deployment should follow empirical risk patterns rather than uniform coverage~\Four{(N=4/9)}.
They cited nightlife districts, dimly lit alleys, and known hot spots during peak hours as priority zones, \quotes{I could see a bunch of robots in Old Town on a Friday night (P2)}, and identified missing-person searches as a high-stakes scenario where MRVS could reduce time-to-locate and officer exposure.
Across these contexts, they articulated a spectrum of desired HRI modes from low-level steering (precise locations or camera angles) to high-level directives such as \quotes{search this area for any anomalies (P7)}, including multi-robot coordination and integration with fixed cameras and drones~\Four{(N=6/9)}.
}
%

\section{Discussion}
\Five{We reflect on our core contributions in relation to prior work, present implications for designing future \system-like systems, and discuss limitations.}
\subsection{\Five{Reflection in Relation to Prior Work}}
\Five{In this section, we situate our contributions within existing research, highlighting how our testbed and system advance prior work across HCI, public-safety, and computer vision.
We reflect on how both the testbed and the \system system were informed by existing research and how they can serve as a foundation for future MRVS studies.}
\subsubsection{\Five{Testbed environment}}
\Five{
Core components in our testbed environment are (1) professional-validated EoI types, Design Requirements of \system-like systems, and (3) a ground-robot video dataset.}

\begin{itemize}
    \item \Five{\textbf{The EoIs Types} (see Table~\ref{tab:EoIfinal}) were informed by prior anomaly detection video datasets that define ``anomaly'' from a computer vision perspective as statistical rarity~\cite{5539872}, researcher-defined categories~\cite{ramachandra2020street}, or labeled from online surveillance footage~\cite{wu2020not}, valuable for benchmarking but not directly actionable in public safety practice.
    We drew EoIs from real campus crime logs (following Clery Act~\cite{cleryact}) on the foundation of existing anomaly datasets~\cite{ma2015anomaly, ramachandra2020street, qian2025ucf, wu2020not, wang2020nwpu, zhang2016single, acsintoae2022ubnormal, pranav2020day, Lu_2013_ICCV, rodrigues2020multi}, then refined with frontline professionals to yield operationally meaningful, context-rich categories (e.g., ``Incident Exposure'').
    This approach grounds ``what to detect'' in professionals' operational realities, translating their priorities into actionable EoIs for future system and model design.
    However, our EoIs remain campus-centric, reflecting a limited jurisdictional perspective and excluding non-visual or hard-to-simulate incidents.
    We therefore position it not as a universal ontology, but as a reusable approach: \system-like systems should expose EoIs as configurable, locally governed ``recipes'' rather than fixed, one-size-fits-all classes.
    The EoI types serve as a reusable template that other jurisdictions can adapt, extend to additional modalities (e.g., audio, text), and use to drive MRVS-like systems and upstream workflows (e.g., dispatch~\cite{liu2023discovering, terrell2004exploring, munro2025video}).
    }
    \item \Five{\textbf{Design Requirements} derived from S1 were informed by HCI’s formative–summative methodological approach, which emphasizes human-centered design for specialized professionals~\cite{leake2024chunkyedit, ganguly2024shadowmagic, yan2022flatmagic}. 
    Our methodological choice shows that domain stakeholders must remain central in defining what systems attend to, why, and under what constraints~\cite{hong2020human, gao2024going, gao2022aligning}.
    The derived design requirements offer guidance for creating \system-like systems that incorporate one or more components of multi-robot operation, video analysis, and decision-making support under attention and time constraints.
    While we instantiated these requirements in particular ways within our implementation, final designs may differ depending on designers’ insights and local operational constraints.
    These requirements also generalize to applications that share characteristics with our user groups, including other high-stakes experts and broader multi-video sensemaking scenarios.}
    \item \Five{\textbf{The Video Dataset} presents ground robot-captured, EoIs-driven, actor-performed videos in campus environments. 
    It complements existing anomaly video datasets which predominantly use short, event-oriented clips recorded by fixed surveillance cameras~\cite{zhu2024advancing} or generated in virtual scenes~\cite{acsintoae2022ubnormal, Sultani_2018_CVPR}.}
    \Three{Following established practice in computer vision and video anomaly detection, our actor-performed simulation approximates rare, safety-critical events that are difficult or unethical to capture opportunistically, similar to canonical datasets that deliberately record scripted or simulated activities in real environments~\cite{blunsden2010behave,mehran2009abnormal,Lu_2013_ICCV}.
    For example, BEHAVE~\cite{blunsden2010behave} provides multi-person interaction videos with actors performing chasing and fighting scenarios, the UMN crowd-panic dataset~\cite{mehran2009abnormal} contains staged escape events, and the CUHK Avenue benchmark~\cite{Lu_2013_ICCV} records abnormal behaviors on a university campus. 
    These actor-driven benchmarks have been used extensively to develop methods that generalize to more realistic ``in-the-wild'' surveillance data, including large-scale real-world anomaly datasets such as UCF-Crime~\cite{Sultani_2018_CVPR}.
    More broadly, work in human action recognition and synthetic video generation (e.g., UCF101, PHAV) shows that models trained on actor-performed or procedurally generated videos improve performance on realistic benchmarks~\cite{soomro2012ucf101,souza2017phav}, indicating that controlled yet diverse scenarios can yield transferable representations.
    Our dataset follows the same pattern: actor-scripted EoIs, co-designed with a police co-author, are embedded in an operational campus environment with uncontrolled bystanders, vehicles, lighting, and occlusions.}
This approach balances ecological validity with experimental control while ensuring coverage of the defined EoIs.
    
\end{itemize}
\subsubsection{\Five{MRVS System Architecture}}
\Five{
As a manifestation of the testbed environment, \system integrates reusable modules be utilized in future systems for multi-video investigation, ground-robot operations, EoIs detection and decision support, and workflows tailored to public safety professionals.
Many anomaly detection systems treat model-centric outputs (e.g., anomaly scores~\cite{Liuetal2017, Sultani_2018_CVPR}, clip labels~\cite{10203771, 10657732}, or captions~\cite{Chen_2023_CVPR, jiang2024visionlanguagemodelsassistedunsupervised}) as the primary product, leaving triage, evidence gathering, and justification to manual operator work.
By contrast, \system treats these as intermediate signals and centers meta-rich artifacts that make reasoning visible: temporal localization, event descriptions, keyframes, and textual rationales that can be checked against the source footage.
Even with imperfect back-end precision, these artifacts are designed to help operators move from ``a possible event'' to ``a defensible judgment'' with lower cognitive and coordination cost~\cite{ehsan2021expanding}, aligning with high-stakes HCI arguments that accountability depends on legible evidence trails rather than end-to-end automation~\cite{rudin2019stop, cai2019hello}.
\system situates the back-end outputs within an interactive front-end that supports the practical mechanics of professional sensemaking reasoning visible to the human decision-maker.
While our current instantiation centers on ground-robot videos, the architecture is source-agnostic: it operates over time-stamped video with optional geospatial metadata, making the same interaction patterns immediately transferable to existing single- or multi-camera infrastructures CCTV, body-worn, in-car, or drone footage, which is a plausible near-term deployment pathway beyond robot-equipped agencies.
}
\subsection{\Five{Implications for Designing Future \system-like Systems}}
\Five{
Drawing on public safety professionals’ reflections on \system in S1 and S2, we outline five design directions for future MRVS-like systems: adaptive workflow enhancement, specialized application deployment, stronger robotics and vision capabilities, and attention to societal considerations.
}
\subsubsection{\Five{Enhancing MRVS Capabilities Within Public-Safety Scenarios}}
\Five{To advance MRVS within existing public-safety workflows, we suggest future systems should incorporate adaptive features that tailor prioritization, scheduling, and operational support to each agency’s unique context and EoI needs.
EoIs can vary substantially by jurisdiction agencies: an activity that is concerning in one city may be routine in another, and situational factors (e.g., crowd density, time of day, weather, and season) shape both incident likelihood and what officers consider noteworthy.
EoIs also vary by the location as not distributed uniformly, as public-safety concerns cluster unevenly across space, with certain neighborhoods or recurring incidents concentrating risk.
A practical direction is therefore to make ``\textit{plug-and-play}'' EoIs configurable, enabling agencies to fine-tune priorities without requiring technical expertise.
More broadly, an adaptive MRVS agent capable of contextualizing data in real time could yield better decision-making outcomes and more relevant alerts for officers in the field.
}
\subsubsection{\Five{Beyond Patrol: Deploying \system Across Specialized Context}}
\Five{
We envision that MRVS can be extended to a wide range of real-world scenarios beyond event detection, supporting specialized, high-stakes operations such as missing-person search, night patrol, and disaster recovery.
These scenarios shift design goals from continuous monitoring with specific EoIs toward rapid coverage, time-to-locate, and resilience under degraded conditions (e.g., low light, debris, unstable connectivity).
Designing for such deployments implies MRVS-like systems should be rugged, quickly deployable, and configurable to mission-specific EoIs, while offering tools for flexible path planning, on-the-fly reprioritization, and collaboration with human teams operating under stress and uncertainty.
}
\subsubsection{\Five{Advancing Core Robotic Foundations for MRVS}}
\Five{
We identify key robotics side implications for advancing MRVS-like systems, including providing flexible control, multi-robot coordination, and robust field-ready platforms to ensure operability and effectiveness.
Multiple abstraction control levels from precise waypoints or camera angles to high-level directives~\cite{porfirio2025uncertainty} such as ``search this area for anomalies'' and supports ``condition-triggered'' handoffs between autonomous and manual modes for both individual robots and the fleet.
Participants anticipated scaling to multi-robot collaboration, including integration with existing surveillance infrastructure and coordination with other robotic platforms such as drones.
\system-like systems should dynamically allocate robotic coverage based on real-time and historical data about hot spot activity.
\system-like systems should deliver high-quality video streams in different formats~\cite{wu2024theia, liu2024muv2, zhang2022m5} to match task demands.
}
\Five{
\system-like deployment holds dual preventive value: persistent patrol enables earlier intervention through anticipatory evidence collection and deters criminal activity through visible presence.
Such moving robotics systems could also help surface deteriorating infrastructure, recurrent hot spots, and patterns of suspicious loitering before they escalate.
Yet sustained preventive deployments raise operational requirements: participants worried about vandalism in areas with heightened anti-police sentiment and about software-level tampering that could compromise system integrity.
They therefore underscored tamper-resistant design, protective mechanisms for critical components, robust physical infrastructure, and minimizing ongoing human intervention during routine operations.
}
\subsubsection{\Five{Advancing Computer Vision Foundations for MRVS}}
\Five{
To better support public-safety workflows, we highlight computer-vision directions that foreground investigator semantics, quantify uncertainty, and address mobile ground-robot video constraints.
\textit{Descriptor-based search} should reason about attributes and object relations when doing retrieval, not just rank by pure visual similarity. 
Pure embedding retrieval (e.g.~SigLIP-style \cite{tschannen2025siglip}) is efficient, but brittle for our application, where dominant colors overwhelmed semantics in these settings, but MLLMs produced attribute descriptions that better matched officer intent. 
Similarly, existing methods do not quantify the confidence of matches well.
New embedding techniques that encode richer attribute–object relations, capture uncertainty, and remain robust under illumination and low-light variation (as surfaced in S2) are needed.}
\Five{
\textit{Re-identification research} likewise needs to move beyond its current emphasis on people and vehicles~\cite{Lu2025CLIPSENetCS, Wang2024SAMdrivenMP}. 
In practice, investigators often need to find specific objects across hours of heterogeneous footage, where identity cues can be mutable or context-dependent (e.g., a convertible with its top open versus closed) and sensing conditions vary widely.
New re-ID capabilities must therefore prioritize investigator-relevant semantics, explicitly handle mutability versus invariance, and remain robust across viewpoint, illumination, and camera-quality shifts.
One possible direction is to leverage MLLM's zero/few-shot capabilities to make embeddings robust with respect to these semantics.}

\Five{
Finally, \textit{robust anomaly detection} for ground-robot video must address peripheral and brief events.
In S2, many missed events were small, short, poorly lit, or near frame edges, suggesting that models trained on centered, sustained actions allocate insufficient attention to the periphery where ground-robot anomalies frequently occur.
Beyond data augmentation, model intervention strategies such as attention intervention~\cite{Chen2025AttentionHD, Li2023InferenceTimeIE, Zhang2023TellYM} and steering~\cite{Wang2024CogSteerCS, Subramani2022ExtractingLS} offer practical routes to increase sensitivity to subtle events.
Additional extensions include active perception and tool usage (e.g., models choosing to zoom).
To reduce false positives without overwhelming operators, agentic loops triaging low-confidence anomaly proposals can identify candidates for secondary verification and refinement.
}
\subsubsection{\Five{Building Responsible and Trustworthy MRVS for Society}}
\Five{
Drawing from our study, we suggest that long-term adoption of MRVS requires embedding ethical and accountable design into core functionality, particularly transparency, reliability, and operability in high-stakes contexts.
Participants stressed that the operational benefits of MRVS-like systems are inseparable from governance.
They valued preventive capabilities, yet warned that many EoIs (e.g., ``hanging around'' or ``loitering'') are socially and historically charged, and uncritical use risks reinforcing over-policing of marginalized groups~\cite{brayne2017big,veale2018fairness}.
Scholarship on predictive policing and big-data surveillance likewise warns that concentrating sensing resources in ``hot spots'' can create feedback loops in which historically over-policed communities are subject to even more intensive monitoring, regardless of actual harm levels~\cite{brayne2017big,petty2018reclaiming}.
These concerns suggest that MRVS-like systems must make what is monitored, where, and under what escalation criteria explicit and contestable, rather than treating deployment as a purely technical optimization problem.}

\Five{
Privacy and public acceptance emerged as context-dependent constraints on sustained deployment.
Participants cautioned that robots may feel intrusive without a clear protective purpose, but are welcomed for targeted tasks like inspections or missing-person searches, and argued that privacy protections should vary by task.
Beyond individual expectations, responsible deployment must address group-level harms, including who is most likely to be continuously observed, how long footage is retained, and whether deployment patterns reinforce existing inequalities~\cite{brayne2017big,whitney2021hci}.}

\Five{
Our results highlighted a practical accountability tension: AI assistance is valuable under time pressure, yet officers remain responsible for decisions and must be able to justify actions taken (or not) in response to system alerts.
To support accountable policing, \system-like tools should embody explainable and transparent AI by making recommendations easy to inspect, contest, and document.
Interfaces should provide brief, human-readable rationales for why clips are flagged, plus fast controls to validate, dismiss, or override suggestions without re-watching long segments.
At the deployment level, MRVS should make its monitoring logic, data retention policies, and limitations legible to agencies and communities, with safeguards against biased over-policing across groups.
}
\subsection{\Five{Limitations and Future Work}}
\Five{We note several limitations that should be considered when interpreting this work.
First, our testbed relies on a simulated multi-robot patrol setting and actor-performed EoIs rather than organically occurring events captured by deployed patrol robots.
Although this actor-performed approach follows established practice in anomaly-detection datasets, it may under-represent edge cases, adversarial behavior, or longer-term behavioral adaptation to robot presence.}
Second, the robots in our study were teleoperated, and video was pre-recorded, so we did not evaluate end-to-end autonomy, navigation failures, or how officers would share attention between live robot control and MRVS interfaces.
Third, our participants were drawn from a limited set of agencies in one U.S. state and evaluated a single campus environment during one season, limiting the generalizability of our EoI taxonomy and design requirements.
Fourth, our evaluation examined the integrated \system experience rather than isolating the causal impact of interface components; future work could employ controlled, feature-by-feature comparisons to understand specific feature effectiveness under different tasks.

\Five{Despite these limitations, the testbed environment and prototype offer a reusable scaffold for future MRVS research.
Near-term deployments may layer \system-like interfaces on existing CCTV, in-car, body-worn, or drone footage in jurisdictions where patrol robots are not yet available.
Future work should also expand EoIs and data collection to additional jurisdictions, seasons, and modalities (e.g., audio, text), and examine long-term adaptation as MRVS systems transition from lab to operational deployment.
}
\section{Conclusion}
In this work, we present \system, a human-AI collaborative system for multi-robot video sensemaking that addresses both practical and technical challenges of integrating ground robot footage into public safety workflows.
By combining insights from public safety professionals with advances in computer vision and robotics, we contribute a real-world testbed environment and a novel system that demonstrates how AI-driven video analysis enables scalable situational awareness with ground robot footage.
Our goal is to advance intellectual contributions for researchers in HCI, computer vision, and HRI who seek to develop impactful, socially grounded AI systems for public safety professionals.
With continued research, this work also aims to provide practical and meaningful capabilities for public safety professionals pursuing more efficient, scalable, data-driven operations—and, ultimately, for citizens who benefit from safer, more resilient communities.
\begin{acks}
This work was supported by the National Science Foundation (NSF2128867, NSF2350352), the Army Research Office (W911NF2320004, W911NF2520011), Google DeepMind (GDM), Clearpath Robotics, FrodoBots Lab, Raytheon Technologies (RTX), Tangenta, the Mason Innovation Exchange (MIX), and Walmart. We thank the George Mason University undergraduate students for their help with video dataset collection. We are grateful to the Northern Virginia police departments for their support and domain guidance, especially Detective Cooper Knight, Major Bull Hudson, Lieutenant Mohammed Tabibi, and Captain Andrew E. Hawkins, as well as other colleagues who provided feedback and guidance. We used generative AI tools for proofreading and editing under full human supervision, in line with the GAIDeT taxonomy (2025)~\cite{suchikova2025gaidet}. The authors are solely responsible for the final manuscript and outcomes.
\end{acks}
\bibliographystyle{ACM-Reference-Format}
\bibliography{99_ref}
\end{document}